# Effect of interface dynamic deformations on instabilities of buoyancy – thermocapillary convection in a two-fluid two-layer system

Alexander Gelfgat

*School of Mechanical Engineering, Faculty of Engineering, Tel Aviv University, Tel Aviv 6997801, Israel*

Abstract

Effect of interfacial disturbances on instabilities of buoyant/thermocapillary convective flows in rectangular cavities is studied in a series of numerical experiments. The computations are carried out for several two-liquid two-layer systems taking into account properties of liquids used in previously published experiments. Relation between the interface deformations and the Boussinesq approximation is discussed. It is shown that in some systems, including the interface disturbances in the model can alter the critical temperature difference by approximately 10%, producing either destabilizing, or stabilizing effect. The interface oscillations appear as standing or travelling waves whose wavelength can vary from short wave lengths to a single wave occupying all the available space. Rough estimations show that in some liquid-liquid systems the interface oscillations amplitude can reach several tens of microns. Patterns of the most unstable disturbances are presented and discussed. It is argued that instabilities in some two layer systems develop similarly to the Holmboe instabilities in stratified mixing layers.

## 1. Introduction

This study addresses the question of importance of interface disturbances in stability problems of combined buoyancy/thermocapillary convection in infinite layers and confined cavities. When single-phase flows with an upper boundary opened to surrounding are studied, the boundary usually is assumed to be flat, and its possible disturbances are usually neglected (see, e.g., Scriven & Sternling, 1964; Zeren & Reynolds, 1972; Afrid & Zebib, 1990; Ben Hadid & Roux, 1992; Parnetier et al. 1993; Mundrane & Zebib, 1994; Gelfgat et al., 1997; Gelfgat, 2007; Lappa, 2009, and references therein). Smith & Davis (1983) included interface disturbances in their study of instability of a single fluid layer and showed that disturbances growth due to interface waves can become most unstable at small Prandtl numbers. Renardy & Renardy (1988) showed that in the absence of Marangoni effect, the interface deformations stabilize the two-layer Rayleigh-Bénard configuration, and replace the monotonic instability by an oscillatory one. Later, McLelland (1995) studied convection in a cavity with free upper surface and reported that for a low Prandtl number fluid ($Pr = 0.015$) the amplitude of the interface oscillations can reach 3-5% of the cavity height. Chen & Hwu (1993) and Mundrane et al. (1995) considered similar problems for fluid with $Pr = 0.01$ and reported smaller oscillations amplitudes that were about 1% of the cavity height.

In more complicated two-phase systems, the interface separating two immiscible fluids is usually assumed to be flat for the base flow state. Its possible disturbances are taken into account when a sole thermocapillary convection is studied (e.g., Vila et al., 1988; Golovin et al., 1995; Hamed & Floryan, 2000; Birikh and Bushueva, 2001; Madruga et al., 2003; Nepomnyashchy et al., 2006; Simanovskii et al, 2012; Vanhaelen, 2019; Patne et al., 2020), while for buoyancy or combined buoyancy/thermocapillary convective flows the interface deformations sometimes are neglected (e.g., Ben Hadid & Roux, 1992; Liu et al., 1993; Wang & Kawahita, 1996, 1998; Nepomnyashchy & Simanovskii, 2006; Liu et al., 2006; Kuhlmann & Albensoeder, 2008), and sometimes are taken into account (Castillo & Velarde, 1982; Wahal & Bose, 1988; Reasenta et al., 1989; Pérez-Garcia & Carneiro, 1990; Regnier et al., 2000; Lebon et al., 2001; Saenz et al., 2013). For example, taking the interface deformations into consideration for the infinite two-layers model, Saenz et al. (2013) predicted interface oscillations in an oil – gas system to be of the order of $10^{-2}$ of the oil layer height.

In several experimental studies, where oscillatory instability of buoyant-thermocapillary convection in cavities was observed (Villers & Platten, 1992; Riley & Neitzel, 1998) no any visible interface deformations were reported. At the same time in earlier experiments of Cerisier et al. (1984) interface oscillations with the amplitude of several microns, when the layer depth was several millimeters, were reported. Later, the interface oscillations with the amplitude above $10\mu m$ were observed in the experiments of Burguete et al. (2001) on 0.65 cSt silicone oil, and, independently, by research group of Duan and Kang, who examined thin layers of Fluoinert FC70 (Kang et al., 2004), 1000 cSt silicone oil (Duan et al., 2006), and 100 cSt silicone oil (Zhang et al., 2014). The interface oscillations were observed when the temperature difference between the hot and cold walls was below 10°C in the experiments of Buruete et al. (2001), and between 10°C and 30°C in the studies of Duang & Kang group. The heat expansion coefficient of these liquids is approximately $10^{-3}\frac{1}{°C}$, so that applicability of the Boussinesq approximation for the temperature differences above 10°C may already be questionable. This observation is in line with several studies that tried to justify neglecting the interface disturbances (Lyubimov et al., 1998; Velarde et al, 2001; Regnier et al., 2000; Lebon et al., 2001; Nepomnyashchy & Simanovskii, 2006) by arguing that their effect lies beyond the applicability of the Boussinesq model. We discuss their arguments below and also offer some counter arguments.

A possible importance of the interface disturbances can be examined by carrying out a simple numerical experiment, in which the interface disturbances are included in the model and then excluded. This was done in the cited above studies of Smith & Davis (1983) and Rosenat et al. (1989), however the author could not find any other research dealing with this question. It should be mentioned also that several studies of the effect of the interface deformations on the onset of primary instability of the motionless state in the Rayleigh-Bénard configuration concluded that these deformations change or can change results of stability studies (see e.g., Zeren & Reynolds, 1972; Wahal & Bose, 1988; Pérez-Garcia & Carneiro, 1991; Regnier et al, 2000; Lebon et al., 2001).

In this study we recall the Boussinesq approximation (see Zeytonian (2003) and references therein) assuming two-fluid two-layer system. We argue that assumptions of a flat interface and exclusion or inclusion of the interface deformations are not connected with the concept of the Boussinesq approximation, which only requires that the maximal temperature difference in the system will be small enough. Further, we consider a simple two-fluid buoyancy-thermocapillary convection model in a rectangular cavity, and address the primary instability of this flow. The interface perturbations are included or excluded from the model. We consider several configurations with realistic liquids and show several examples, in which interface disturbances have only negligible effect, and others, where they can noticeably change the critical parameters. Further, we recall that interface waves, even those with very small amplitude, can be detected by non-intrusive experimental means (Kang et al., 2004; Duan et al., 2006; Zhang et al., 2014; Nezhihovski et al., 2022), which yields quite an effective tool for experimental measurement of such instabilities. Making an assumption of possible velocity perturbation amplitude, we estimate amplitude of the interfacial waves. Finally, we discuss the observed flow patterns and patterns of the most unstable disturbances. A particularly interesting finding here is an observed similarity between instability onset in some of the considered systems with the Holmboe instability taking place in stratified mixing layers.

## 2. Formulation of the problem
### 2.1. General equations

We consider combined two-dimensional buoyancy – thermocapillary convection in a system of two immiscible fluids inside a rectangular cavity with the height $H$ and the length $L$. The fluids 1 and 2 occupy, respectively, lower and upper parts of the cavity, so that the average heights of the fluid sublayers are $h$ and $H - h$. The geometry and a possible flow patterns are sketched in Fig. 1. It is assumed that in a fully developed steady flow, the two immiscible fluids form a two-layer system, making no assumptions yet about shape of the interface. Both fluids are assumed to be incompressible when they are isothermal. As will be shown below, at some parameters the flow consists of clockwise and counter clockwise circulations as is shown in Fig. 1a, but in other cases two clockwise circulations are separated by a counter clockwise one, which can be located either in the upper (Fig. 1b), or in the lower (Fig. 1c) layer (see also Villers & Platten, 1990). The applicability of the Boussinesq approximation for this case was discussed by Mihaljan (1962), who showed that two dimensionless parameters discussed below must be small.

The dimensionless flow in the Boussinesq approximation is described by velocity $\boldsymbol{u}_j = (u_j, w_j)$, $j = 1$ or 2, of the liquids 1 and 2, governed by the momentum and continuity equations defined into each liquid. The dynamic pressure $p_j$ in both liquids is replaced by $P_j = p_j + \rho_j g z$, which combines the dynamic and hydrostatic pressure. Assuming $0 \leq x \leq L$ and $0 \leq z \leq H$, the governing equations and the boundary conditions read

$$\rho_j \left[\frac{\partial \boldsymbol{u}_j}{\partial t} + (\boldsymbol{u}_j \cdot \nabla)\boldsymbol{u}_j\right] = -\nabla P_j + \mu_j \Delta \boldsymbol{u}_j + \rho_j g \boldsymbol{e}_z \quad , \tag{1}$$

$$\nabla \cdot \boldsymbol{u}_j = 0, \tag{2}$$

$$\rho_j C_{p,j} \left[\frac{\partial T_j}{\partial t} + (\boldsymbol{u}_j \cdot \nabla) T_j\right] = k_j \Delta T_j \tag{3}$$

with no-slip boundary conditions at the cavity walls

$$\boldsymbol{u}_j(x=0,z) = \boldsymbol{u}_j(x=L,z) = \boldsymbol{u}_j(x,z=0) = \boldsymbol{u}_j(x,z=H) = 0, \tag{4}$$

and continuity of velocities and stresses at the interface

$$\boldsymbol{n} \cdot (\boldsymbol{T}_1 \cdot \boldsymbol{n}) = \boldsymbol{n} \cdot (\boldsymbol{T}_2 \cdot \boldsymbol{n}), \quad \boldsymbol{\tau} \cdot (\boldsymbol{T}_1 \cdot \boldsymbol{n}) = \boldsymbol{\tau} \cdot (\boldsymbol{T}_2 \cdot \boldsymbol{n}). \tag{5,6}$$

Here $t$ is time, $\rho_j, \mu_j, C_{p,j}$, and $k_j$ are density, viscosity, heat capacity, and thermal conductivity of the fluids, respectively, $T_j$ is the viscous stress tensor, $\boldsymbol{n}$ is the unit normal to the interface, and $\boldsymbol{\tau}$ is the unit vector tangent to the interface. Boundary conditions for the temperature will be specified later. In the following we assume that the densities $\rho_j$ and the interface surface tension $\sigma$ decay linearly with the temperature and do not depend on any other parameters, e.g., pressure. This yields

$$\rho_j = \rho_{j,0}[1 - \beta_j(T - T_0)], \quad \sigma = \sigma_0 - \gamma(T - T_0). \tag{7,8}$$

Here $\beta_j$ are thermal expansion coefficients of the liquids, and $\gamma = -\left(\frac{d\sigma}{dT}\right)$. According to the analysis of Mihaljan (1962), the Boussinesq approximation is valid if

$$\varepsilon_1 = \beta_j(T_{max} - T_{min}) \ll 1, \quad j = 1,2 \tag{9.1}$$

$$\varepsilon_2 = \frac{\alpha_j^2}{c_{p,j} H^2 (T_{max} - T_{min})} \ll 1, \quad j = 1,2 \tag{9.2}$$

Where $T_{max}$ and $T_{min}$ are the maximal and minimal temperatures of the system, respectively, and $\alpha_j = k_j / c_{p,j} \rho_j$ are thermal diffusivities of the liquids. The smallness of the first parameters means that changes of the density are much smaller than the density itself. The smallness of the second parameter allows one to neglect temperature changes due to viscous dissipation and changes of the pressure. For the liquids considered below the thermal diffusivity varies between $10^{-8}$ and $10^{-4} \frac{m^2}{s}$, and the heat capacity is of the order of $10^3 \frac{J}{kg \, °C}$. Therefore, the parameter $\varepsilon_2$ remains small for all reasonable values of the cavity height and the temperature difference. It should be underlined that the Boussinesq approximation treats only temperature dependence of the density. Within the Boussinesq model, a flow is treated as incompressible, which can be applied not only to convective flows, but to other non-isothermal flows as well (e.g., non-isothermal shear flows, see Gelfgat & Kit (2006) and references therein). Additionally, the Boussinesq model includes the density variations into a non-potential part of the buoyancy force, while its potential part is added to the pressure (Landau & Lifshitz, 1987). Temperature dependences of other parameters, including the surface tension, do not relate to the Boussinesq approximation and can be included in the Boussinesq, as well as in non-Boussinesq models.

Thus, interpreting a temperature dependent viscosity as a non-Boussinesq effect, as was done by Velarde et al. (2001), is inconsistent with the derivation of the Boussinesq approximation (see, e.g., Mihaljan, 1962).

Choosing $H$, $\mu_1/\rho_1 H$, $H^2\rho_1/\mu_1$, and $\mu_1^2/H^2\rho_1$ to be the scales of length, velocity, time, and pressure, respectively, rendering the temperature dimensionless by $\theta = (T - T_{min})/(T_{max} - T_{min})$, and assuming the Boussinesq approximation justified by Eq. (9), we arrive to the non-dimensional form of equations

$$\left[\frac{\partial \boldsymbol{u}_j}{\partial t} + (\boldsymbol{u}_j \cdot \nabla)\boldsymbol{u}_j\right] = -\frac{\rho_{2,0}}{\rho_{21}\rho_{j,0}}\nabla P_j + \mu_j \frac{\rho_{2,0}}{\rho_{21}\rho_{j,0}} \frac{\mu_{21}\mu_j}{\mu_2} \Delta \boldsymbol{u}_j + Gr \frac{\beta_{21}\beta_j}{\beta_2} \theta_j \boldsymbol{e}_z, \tag{10}$$

$$\nabla \cdot \boldsymbol{u}_j = 0, \tag{11}$$

$$\frac{\partial T_j}{\partial t} + (\boldsymbol{u}_j \cdot \nabla)\theta_j = \frac{\alpha_{21}\alpha_j}{\alpha_2}\frac{1}{Pr}\Delta\theta_j, \tag{12}$$

where the dimensionless governing parameters are the density, viscosity, thermal expansion, and thermal diffusivity ratios $\rho_{21} = \rho_2/\rho_1$, $\mu_{21} = \mu_2/\mu_1$, $\beta_{21} = \beta_2/\beta_1$, and $\alpha_{21} = \alpha_2/\alpha_1$ ($\alpha_j = k_j/\rho_j C_{p,j}$), the Prandtl number $Pr = \mu_1/\alpha_1\rho_1$, and the Grashof number $Gr = g\beta_1(T_{max} - T_{min})H^3\rho_{1,0}^2/\mu_1^2$. Note, that equations for the fluid 1 are the same as they would be for a single phase buoyancy convection flow, while ratios of thermophysical properties necessarily appear in the equations describing the flow in the fluid 2. The stream function $\psi_j$ is defined separately for each fluid layer as $u_j = -\partial\psi_j/\partial z$, $w_j = \partial\psi_j/\partial x$, and is used only for visualization purposes.

For further derivation of the boundary conditions we assume that the interface deviates from the level $z = h$, so that its dimensionless position is given by

$$z_{interface}(x,t) = \hat{h} + \chi(x,t), \quad \hat{h} = \frac{h}{H} \tag{13}$$

The boundary conditions at the interface for the normal and tangent stresses read (Landau & Lifshitz, 1987; Nepomnyashchy et al., 2006)

$$p_1 - p_2 - Ga\left(1 - \frac{1}{\rho_{21}}\right)\chi - \frac{Ga}{Bo}K = -\left[\frac{\partial u_{2,1}}{\partial x_k} + \frac{\partial u_{2,1}}{\partial x_i} - \mu_{21}\left(\frac{\partial u_{2,2}}{\partial x_k} + \frac{\partial u_{2,2}}{\partial x_i}\right)\right]n_i n_k, \tag{14}$$

$$\left[\frac{\partial u_{1,i}}{\partial x_k} + \frac{\partial u_{1,k}}{\partial x_i} - \mu_{21}\left(\frac{\partial u_{2,i}}{\partial x_k} + \frac{\partial u_{2,k}}{\partial x_i}\right)\right]n_k - Mn\frac{\partial \theta}{\partial x_i}\tau_i = 0 \quad, \tag{15}$$

where $K$ is the dimensionless main interface curvature, $\boldsymbol{n}$ and $\boldsymbol{\tau}$ are unit normal and tangent to the interface vectors, respectively, $Ga = \frac{g\rho_{1,0}^2 H^3}{\mu_1^2}$ is the Galileo number, $Bo = \rho_{1,0}gH^2/\sigma$ is the Bond number, $Ma = \frac{\gamma(T_{max}-T_{min})H}{\mu_1\alpha_1}$ is the Marangoni number, and $Mn = Ma/Pr$.

Additional boundary conditions at the interface are continuity of the velocity, the temperature, and the heat flux

$$\boldsymbol{u}_1 = \boldsymbol{u}_2, \quad \theta_1 = \theta_2, \quad \left(\frac{\partial \theta_1}{\partial x_i} - k_{21}\frac{\partial \theta_2}{\partial x_i}\right)n_i = 0, \tag{16}$$

and the kinematic condition

$$w = \frac{\partial \chi}{\partial t} + u\frac{\partial \chi}{\partial x}. \tag{17}$$

*2.2. Interface deformations and Boussinesq approximation*

As mentioned in the Introduction, several authors (Lyubimov et al., 1998; Velarde et al, 2001; Renier et al., 2000; Lebon et al., 2001; Nepomnyashchy & Simanovskii, 2006) argue that if the Galileo number is large, $Ga \gg 1$, then the effect of the interface deformation is non-Boussinesq, so that it should be neglected if $\rho_{21}$ is far from unity. Note, that according to the main assumption of the Boussinesq model, $Gr/Ga = \beta_j(T_{max} - T_{min}) \ll 1$, Eq. (9.1). Thus, $Ga \gg Gr$, and for the terrestrial conditions the Grashof number usually exceeds $10^3$. This means that if the condition (9) holds, then $Ga \gg 1$, so that no additional assumptions about the Galileo number are needed. On the other hand, if the Galileo number is much larger than the Grashof number, and therefore much larger not only than unity, but also than, say, $10^3$, all the terms of Eq. (14) that do not contain the Galileo number seemingly can be dropped as small, which will lead to a necessary conclusion $\chi = 0$.

A more rigorous argument follows from an asymptotic expansion into the series of $\varepsilon_1$ (Nepomnyashchy et al., 2006). It can be shown that in the leading order $\chi \sim \varepsilon_1/Ga$, so that $\chi \ll \varepsilon_1$ and, consequently, the interface disturbances can be neglected for small $\varepsilon_1$, i.e., within the Boussinesq approximation. This is followed by the next conclusion that the interface disturbances are a non-Boussinesq effect, and should not be included into the Boussinesq model.

We can counter argue that if the interface deformation is neglected, then the assumption $\chi = 0$ necessarily means $K = 0$, so that both terms of Eq. (14) proportional to the Galileo number vanish. On the other hand, assuming that the amplitude of the interface deformation is estimated as $\varepsilon_1/Ga$, as one would expect in the terrestrial conditions, the two terms of Eq. (14) containing $\chi$ are product of a large Galileo number by a small interface deformation, or a small interface curvature. Moreover, the magnitude of this product is estimated as $\varepsilon_1$, so that, within the mentioned asymptotic expansion, they become comparable with the other terms of Eq. (14). Therefore, a large Galileo number means that the interface deformations are expected to be very small, however, their contribution to the balance of normal stresses can be non-negligible. Seemingly, this was the case reported by Renardy & Renardy (1988), where interface deformations exhibited a stabilizing effect, while the deformations themselves remained small.

In fact, in any finite container the capillary interface forms so-called wetting angles at the boundaries. This means that it can be mathematically flat only if there is no lateral boundaries, i.e., for an infinite layer, or if the wetting angle $\varphi_{wetting}$ is exactly 90°, which is, of course a quite exceptional case. For $\varphi_{wetting} \neq 90°$, the interface shape is defined by the equation (Landau & Lifshitz, 1987)

$$p_1 - p_2 + Ga\left(1 - \frac{1}{\rho_{21}}\right)\chi - \frac{Ga}{Bo}K + \tau_{n,2} - \tau_{n,1} = 0, \tag{18.1}$$

with the wetting angle boundary conditions, and the conservation of volume, which in the present notations read

$$\left(\frac{\partial \chi}{\partial x}\right)_{wall} = cotan(\varphi_{wetting}), \quad \int_0^L \chi(x)dx = 0, \tag{18.2}$$

where $\tau_{n,2}$ and $\tau_{n,1}$ are normal viscous stresses at both sides in the interface. In the motionless fluid Eq. (18.1) reduces to the well-known the Laplace-Young equation. Note that a non-horizontal shape of the actual interface, defined by the problem (18), was demonstrated experimentally by Li et al. (2014), and was modelled numerically by, e.g., Hamed & Floryan (2000) and Fedyushkin (2020). Apparently, in the terrestrial conditions, the resulting function $\chi(x)$ only slightly deviates from zero, but is not an exact zero. This returns us to the above counter argument.

Another counter argument arises when linear stability of some known base flow is studied. Assume that the pressure, velocity and interface deformation in Eq. (14) are infinitesimally small disturbances, for which the asymptotic expansion in the power series of $\varepsilon_1$ can be invalid. Then one can rise the following question: can very small disturbances of $\chi$, being multiplied by a large Galileo number, become large enough to affect other terms of Eq. (14)? If this is possible, then perturbations of the interface should be included in the linearized stability problem, even if initial interface deformations were neglected when the base flow was derived.

To strengthen the above argument, we refer to our sketch in Fig. 1a. Consider a case of buoyancy convection ($Mn = 0$) when both layers are heated from one side and are cooled from another, as is shown in the figure. Both liquids tend to ascend along the hot cavity wall descent along the cold one, thus forming two convective circulations. Note that in this case the flow direction of one fluid along the interface tends to be opposite to the flow direction of another one, as is shown schematically in Fig 1a. Thus, in the middle of the cavity, there can exist a mixing layer configuration. It is well-known that in the inviscid limit mixing layers are unstable owing to the Kelvin-Helmholtz instability mechanism (Chandrasekhar, 1961), while their critical Reynolds number within the viscous model is quite small (Gelfgat & Kit, 2006). It is well-known also that derivation of the Kevin-Helmholtz instability for a two-layer system involves perturbations of the interface (Chandrasekhar, 1961). Even more extreme example can be given by placing a heavier liquid above a lighter one. If both liquids are non-uniformly heated, one may wish to apply the Boussinesq approximation. Apparently, such a configuration will be subject to the Rayleigh-Taylor instability, which also involves interface disturbances (Chandrasekhar, 1961), and does not alter the Boussinesq approximation in any way.

*2.3. Linear stability problem*

In the following we perform numerical study of linear stability of a two-dimensional two-layer flow in a rectangular cavity. The flow is driven by the buoyancy and thermocapillary forces. We assume that the surface tension at the liquid-liquid interface is large and that wetting angle is exactly 90°, so that in motionless system the solution of the problem (18) results in the perfectly flat interface. Alternatively, we can recall the experimental configuration described in Shatz & Neitzel (2001), where the contact line was pinned at an edge at the brim of the cavity sidewall. The interface will remain perfectly flat in the presence of flow, if the pressure at the two sides of the interface will be equal. The latter is usually assumed for two phase parallel flows in pipes and channels (see Gelfgat & Brauner (2020) and references therein), and we add this assumption to the present flow model. Thus, the assumption that $\chi \approx 0$ owing to $Ga \gg Gr \gg 1$ is not formally involved in evaluation of the base flow. At the same time, recalling the above discussion, we take into account infinitesimal disturbances of the flat interface of the base flow.

The finite length of the cavity allows us to exclude the long wave instability, for which accounting for the interface disturbances can be problematic (Lyubimov et al., 1998; Velarde et al., 2001). At the same time, the aspect ratio of the cavity is chosen large enough, $A = 5$, to allow for development of interfacial waves.

Assuming that $\boldsymbol{U}_j = (U_j, W_j), P_j$, and $\theta_j$, $j = 1,2$, are solution of the steady problem (1)-(17) with a flat liquid-liquid interface, and defining the infinitesimally small disturbances of velocity, pressure, temperature and interface as $[\widetilde{\boldsymbol{v}}_j(x,z), \tilde{p}_j(x,z), \tilde{\theta}_j(x,z), \chi(x)]exp(\lambda t)$, $\widetilde{\boldsymbol{v}}_j = (\tilde{u}_j, \widetilde{w}_j)$, the linearized stability problem reads (for details see Gelfgat & Brauner (2020))

$$\lambda \widetilde{\boldsymbol{v}}_j + (\boldsymbol{U}_j \cdot \nabla)\widetilde{\boldsymbol{v}}_j + (\widetilde{\boldsymbol{v}}_j \cdot \nabla)\boldsymbol{U}_j = -\frac{\rho_{2,0}}{\rho_{21}\rho_{j,0}} \nabla \widetilde{p}_j + \mu_j \frac{\rho_{2,0}}{\rho_{21}\rho_{j,0}} \frac{\mu_{21}\mu_j}{\mu_2} \Delta \widetilde{\boldsymbol{v}}_j + Gr \frac{\beta_{21}\beta_j}{\beta_2} \widetilde{\theta}_j \boldsymbol{e}_z ,$$
(19)

$$\nabla \cdot \widetilde{\boldsymbol{v}}_j = 0.$$
(20)

$$\frac{\partial \widetilde{\theta}_j}{\partial t} + (\boldsymbol{U}_j \cdot \nabla)\widetilde{\theta}_j + (\widetilde{\boldsymbol{v}}_j \cdot \nabla)\theta_j = \frac{\alpha_{21}\alpha_j}{\alpha_2} \frac{1}{Pr} \Delta \widetilde{\theta}_j,$$
(21)

$$\widetilde{\boldsymbol{v}}_j(x=0,z) = \widetilde{\boldsymbol{v}}_j(x=A,z) = \widetilde{\boldsymbol{v}}_j(x,z=0) = \widetilde{\boldsymbol{v}}_j(x,z=1) = 0,$$
(22)

$$\widetilde{\theta}_1(x,z=\hat{h}) = \widetilde{\theta}_2(x,z=\hat{h}), \quad \left(\frac{\partial \widetilde{\theta}_1}{\partial x_i} - k_{21} \frac{\widetilde{\theta}_2}{\partial x_i}\right)_{z=\hat{h}} n_i = 0,$$
(23)

$$\lambda \chi = \widetilde{w}_1(x, z=\hat{h}) + U_1(x, z=\hat{h}) \frac{\partial \chi}{\partial x},$$
(24)

$$\widetilde{w}_1(x, z=h_1) = \widetilde{w}_2(x, z=h_1)$$
(25)

$$\widetilde{u}_1(x, z=\hat{h}) = \widetilde{u}_2(x, z=\hat{h}) + [U_2(x, z=\hat{h}) - U_1(x, z=\hat{h})]\chi$$
(26)

$$\left[\widetilde{p}_2 - \widetilde{p}_1 + 2\left(\frac{\partial \widetilde{w}_1}{\partial z} - \mu_{21} \frac{\partial \widetilde{w}_2}{\partial z}\right) - \frac{Ga}{Bo}\left(\frac{\partial^2 \widetilde{\eta}}{\partial y^2} - k^2 \chi\right) - Ga(\rho_{21} - 1)\chi\right]_{z=\hat{h}} = 0$$
(27)

$$\left[\left(\frac{\partial^2 U_1}{\partial z^2} - \mu_{21} \frac{\partial^2 U_2}{\partial z^2}\right)\chi - 2\left(\frac{\partial U_1}{\partial x} - \mu_{21} \frac{\partial U_2}{\partial x}\right)\frac{\partial \chi}{\partial x} + \left(\frac{\partial \widetilde{u}_1}{\partial z} - \mu_{21} \frac{\partial \widetilde{u}_2}{\partial z}\right) + \left(\frac{\partial \widetilde{w}_1}{\partial x} - \mu_{21} \frac{\partial \widetilde{w}_2}{\partial x}\right)\right]_{z=\hat{h}} =$$
$$-Mn \left(\frac{\partial \widetilde{\theta}_1}{\partial x}\right)_{z=\hat{h}}$$
(28)

Equations (19)-(28) define the generalized eigenvalue problem

$$\lambda \mathbf{B}(\widetilde{\boldsymbol{v}}, \widetilde{p}, \widetilde{\theta}, \chi)^T = \mathbf{J}(\widetilde{\boldsymbol{v}}, \widetilde{p}, \widetilde{\theta}, \chi)^T$$
(29)

for the complex time increment $\lambda$ and the eigenvector $(\widetilde{\boldsymbol{v}}, \widetilde{p}, \widetilde{\theta}, \chi)$ being a possible perturbation. The Jacobian matrix **J** is obtained after discretization of the linearized problem (19)-(28). **B** is a diagonal matrix, whose diagonal elements corresponding to the time derivatives of $\widetilde{\boldsymbol{v}}$ and $\widetilde{\theta}$ are equal to one, while the elements corresponding to $\widetilde{p}$ and to the boundary conditions are zeros, so that $det\mathbf{B} = 0$. Thus, the generalized eigenproblem (29) cannot be transformed into a standard one, and the eigenvalue problem is solved in the shift-and-inverse formulation (Gelfgat, 2007).

According to the linear stability theory, the flow is unstable if there exists at least one eigenvalue with a positive real part. In the course of computational process, we monitor the eigenvalue having the largest real part. This eigenvalue and its eigenvector are called leading, or, for the purposes of stability studies, the most unstable. The instability threshold corresponds to a set of governing parameters at which the leading eigenvalue crosses the imaginary axis, so that its real part turns from negative to positive.

3. **Numerical method**

The governing equations were discretized by the finite volume method. The steady flows were calculated by Newton iteration, and the eigenvalues of the linear stability problem were computed by the Arnoldi iteration in the shift-and-inverse mode. The numerical technique is described in Gelfgat (2007), and additions required to account for the interface disturbances are discussed in Gelfgat & Brauner (2020). The convergence was examined separately for steady flows and for the critical numbers. Results of the convergence study are reported in the Supplementary material.

Calculation of steady flows was verified against results of Liu et al. (1993) obtained for $A = 2$. We monitored the convergence of the minimal and maximal values of the stream function, and the Nusselt numbers at the heated boundary, which were calculated separately for the lower and upper layer. Within our numerical approach we arrive to the converged fourth decimal place using 700×700 grid. The results of Liu et al. (1993) were obtained on much coarser 82×161 grid, so that only two first decimal places agree with our results.

Convergence of the critical numbers was studied for five different liquid pairs (see below). The stability studies are usually more computationally demanding, so that using the 2000×300 grid we can ensure convergence of only two decimal places (see Supplementary material), however, this is enough for the purposes of this study.

Note, that within the linear stability analysis, the linearized formulation applied in this work considers only infinitesimal interface deformations, so that no interface tracking is needed.

4. **Numerical results**

To make our results closer to realistic flow configurations, we choose two-fluid systems that already appeared in previous studies. In four first systems the lower liquid is always water, whose properties are taken at $25°C$. The upper liquids are air (Zhao et al., 1995; Fedyushkin, 2020), benzene (Zeren & Reynolds, 1972), 1cSt Dow Corning oil (Zhao et al., 1995), and AK20 silicone oil (Straub et al., 1990; Huang et al., 2014). Additionally, we consider a two-layer system of perfluorinated hydrocarbon FC70 liquid and 3 cSt silicone oil (Nejati et al., 2015).

Properties of all the liquids are listed in Table 1. In the following we consider the cavity aspect ratio $A = 5$, and equal heights of both layers, $\hat{h} = 0.5$.

The results of computational modelling reported below are aimed to present several examples, in which taking the interface disturbances into account within the linear stability problem and in the Boussinesq approximation, either is negligible, or changes the result. We examine also how the thermocapillary flow along the interface affects the base flow and its instability. According to the above discussion about characteristic amplitudes of the interface and velocity disturbances, we expect to observe that the amplitudes of the interface perturbations will be noticeably smaller than that of the velocity field. According to the asymptotic expansion results, their ratio must be of the order $Ga^{-1}$, which also will be verified.

Note, that since both Grashof and Marangoni numbers are proportional to the temperature difference $\Delta T$, they cannot be varied independently in an experiment. Therefore, in the following we consider $\Delta T$ as a meaningful critical parameter. The dimensionless parameters, as well as the dimensional values of $Gr/\Delta T$ and $Mn/\Delta T$ are calculated for $H = 10\ cm$, and are also included in Table 1. For the discussion purposes we always assume the cavity height to be either $1\ cm$ or $10\ cm$, which are a reasonable sizes for a possible experimental setup. Also, we assume that if the layers are not very thin, there is no thin film effects, and behavior of the interface disturbances will be similar. Therefore, the layers are assumed to be of the same heights, i.e., $\hat{h} = 0.5$.

All the instabilities reported below appear to be a transition from steady to oscillatory state, which takes place owing to the Hopf bifurcation. Along with the critical temperature difference, this transition is characterized by the critical frequency of oscillations $\omega_{cr}$, which is yielded by the imaginary part of the dominant eigenvalue that crosses the imaginary axis. The critical frequencies are reported below together with the critical temperature differences.

In the following numerical experiments we assume that the left and right vertical boundaries of the cavity have constant temperatures, $T_{max}$ and $T_{min}$, respectively, while the horizontal boundaries are either perfectly conducting or perfectly insulating. These result in the following dimensionless boundary conditions for the base state temperature and its perturbation

$$\theta(x=0,z) = 1, \quad \theta(x=A,z) = 0, \quad \tilde{\theta}(x=0,A;z) = 0, \tag{30}$$

for perfectly conducting horizontal boundaries

$$\theta(x,z=0,1) = 1 - \frac{x}{A}, \quad \tilde{\theta}(x;z=0,1) = 0, \tag{31}$$

and for perfectly insulated horizontal boundaries

$$\frac{\partial \theta}{\partial z}(z=0,1) = \frac{\partial \tilde{\theta}}{\partial z}(z=0,1) = 0 \ . \tag{32}$$

As mentioned, in the following computations we fix the cavity aspect ratio at $A = 5$, which is assumed to be large enough to allow development of waves along the interface. For each liquids pair we calculate the critical temperature difference for three cases: (i) taking into account interface disturbances, (ii) excluding interface disturbances from the linear stability problem, and (iii) taking into account interface disturbances and excluding the thermocapillary force. All the results are summarized in Tables 2-6. To compare amplitude of the interface disturbances to the amplitude of flow perturbations, we normalize $\chi$ by the maximal value of the horizontal velocity perturbation. The ratios $|\chi_{max}|/|\tilde{u}_{max}|$ are included in Tables 2-5. Additionally, we estimate the dimensional amplitude of the interface disturbances assuming that at some slight supercriticality the amplitude of horizontal velocity perturbation is 1% of the maximal value of the horizontal velocity of the base flow. Thus, the dimensional amplitude is estimated as $H \cdot |\chi_{max}| \cdot (0.01|u_{max}|/|\tilde{u}_{max}|)$, where $H$ is $1\ cm$ or $10\ cm$. These estimations are summarized in Table 6.

*4.1 "Large" cavity, $H = 10\ cm$, conducting horizontal boundaries*

In the first series of numerical experiments we consider a cavity whose height and length are 10 cm and 50 cm, respectively, and the horizontal boundaries are perfectly conducting. The corresponding results are summarized in Table 2.

Our first observation is that the critical temperature differences $\Delta T_{cr}$ for this case, are well below 1°C, and taking into account the thermal expansion coefficients listed in Table 1, we conclude that the condition of the Boussinesq approximation $\beta \Delta T_{cr} \ll 1$ holds for all the systems considered. The second observation is that the thermocapillary force exhibits quite a weak effect on the flow instability, so that for this configuration the buoyancy force is dominant. Also, normalized amplitudes of the interface oscillations are not noticeably affected by the thermocapillary force. The third observation is that taking into account disturbances of the interface, leads to a noticeable change in $\Delta T_{cr}$ for the system with AK 20 oil, while in four other systems no noticeable change is observed. A quite unexpected results is that in the case of AK20 oil – water system the interface disturbances amplitude is several orders of magnitude larger than in three other cases. One can argue that this is because the larger density of the AK20 oil, which is the closest to the density of water, which makes the term with the Galileo number in Eq. (14) smaller.

Estimations of the dimensional amplitude of the interface disturbances (Table 6) show that a noticeable amplitude, as already expected, is observed only in the AK20 oil – water system, where it reaches several tens of microns. The smallest dimensional estimate is observed for the 1 cSt oil – water system (Table 6), for which also the ratio $\frac{|X|_{max}}{|u|_{max}}$ is the smallest. However, results of Table 2 show that including of these small disturbances in the model alters the critical temperature difference by approximately 8%, which is noticeably larger than in the other configurations with water. Another interesting observation is exclusion of the thermocapillary force from the AK20 oil – water model, which increases the interface waves amplitude in more than an order of magnitude. This shows that is some cases the Marangoni effect can suppress interface waves.

Recalling that the theoretical estimation based on the power series of $\varepsilon_1$ predicts $\frac{|X|_{max}}{|u|_{max}} \sim 1/Ga$, we can state that it possibly holds for the first three systems listed in Table 2, but this is definitely not the case for the last two systems of AK20 oil – water and 3cSt oil – FC70. We can speculate that in the first three cases the instability is affected by buoyancy, while in the two last cases is not. This issue is further addressed below.

It is quite apparent that the differences in the critical parameter values and the interface amplitudes result from qualitatively different base flows that become unstable owing to qualitatively different perturbations. To illustrate that we report the base flow patterns in Fig. 2, base flow temperature distributions in Fig. 3, and the perturbations of the stream function in Fig. 4. The flow patterns are reported for the values of $\Delta T$ close to the critical ones. It is clearly seen that the flow and perturbation patterns are qualitatively different.

In Fig. 2 we observe that the two clockwise circulations driven mainly by the buoyancy force are "connected" with a thin reverse counter clockwise circulation. This is a quite common feature of these flows that allows for the same circulation direction in the upper and lower circulations, which are more intensive than one located between them. Note also that depending on fluids properties, the counter clockwise circulation can be located in the upper (Fig. 2a), or in the lower (Fig. 2b-2e) fluid, as is shown schematically in Fig. 1. In the air-water, benzene-water, and 1 cSt oil - water systems the upper circulation is noticeably more intensive, which is an apparent consequence of a smaller viscosity of the upper liquid. When viscosity of the upper liquid is larger, the size and intensity of counter-clockwise circulation decrease, so that it becomes almost unnoticeable in the 3 cSt – FC70 configuration (Fig. 2e).

The isotherms shown in Fig. 3 are stronger curved in the areas where the flow is more intensive. Besides this, we observe temperature boundary layers near the vertical boundaries, in which the temperature variation along the horizontal coordinate becomes very steep. In these boundary layers we see also accumulation of warmer fluid near the lower part of the left (hot) boundary (Figs. 3a and 3e), and accumulation of colder fluid at the upper parts of the right (cold) boundary (Fig. 3b), which creates local unstable stratification. Apparently, the boundary layers themselves, as well as the unstable stratifications can be sources of instability. We observe this in the patterns of the stream function perturbations (Fig. 4), where perturbations are localized near the lower part of the left (hot) boundary (Figs. 4a and 4e), or near the upper part of the right (cold) boundary (Figs. 4b and 4d). The only exception is the 1 cSt oil – water case (Fig. 4c), in which the most unstable disturbance is located in the bulk of the lower liquid. Note that the most unstable disturbances of the two last systems (Figs. 4d and 4e) are localized in the area of boundary layers, so that the instability is possibly caused by a steep change of velocity shear. This supports the above made speculation stating that these instabilities are not directly affected by buoyancy.

Graphs of the interface disturbances are plotted in Fig. 5. The left parts of the figure illustrates convergence of $|\chi|$. The well-established convergence makes us confident that these small-amplitude interface oscillations are not spurious. The right part of the graphs show the real, imaginary and absolute values of $\chi$. Additionally, the interface oscillations are animated in the supplemented file. We observe that in the air – water configuration the interface oscillates with a single amplitude maximum at the interface center. In the three other cases the interface oscillations are wavy. In the cases of benzene – water system the real and imaginary parts oscillate in counter phase, while in the AK20 oil – water system they oscillate in phase. In both cases the interface oscillations result in a standing wave, as is clearly seen in the animation. In the case of 1 cSt oil – water the real and imaginary parts of $\chi$ are not in phase, so that the wave is expected to propagate along the interface. The latter also can be clearly observed in the animation. The interface disturbances in the case of 3 cSt oil – FC70 are similar to those observed in the AK20 oil – water case, with the wavelength approximately 10 times smaller. They are not included in Fig. 5 and in the animation.

*4.2 "Large" cavity, $H = 10\ cm$, insulating horizontal boundaries*

When the perfectly conducting boundaries are replaced by perfectly insulating ones, the flows stabilize, so that critical temperature differences become larger (Table 3). The most striking effect is observed in the air – water system, for which $\Delta T_{cr}$ becomes larger than 50°C, so that the Boussinesq approximation becomes inaccurate. In the other cases we observe that, contrarily to the systems with perfectly conducting boundaries, excluding the thermocapillary force ($Ma = 0$) change the stability results stronger. The same can be said about excluding the interface disturbances ($\chi = 0$). It should be noted, that the flows still are driven mainly by the buoyancy force, so that patterns of base flows remain similar, while the values of critical parameters differ for the full and partial models.

Looking at the base flow streamlines (Fig. 6) we observe that the weak counter clockwise circulations become very thin, while the two clockwise circulations become more intensive and form boundary layers near the vertical boundaries. These boundary layers develop together with thinner temperature boundary layers that are observed in isotherms plotted in Fig. 7. Such boundary layers usually develop in cavities with perfectly insulating (or adiabatic) horizontal boundaries and were intensively discussed in literature (see Lappa (2009) and references therein). All the perturbation patterns reported in Fig. 8 show that the instabilities develop inside the boundary layer regions, sometimes near the hot border (Fig. 8a and 8b), sometimes near the cold one (Fig. 8d), or near both (Fig. 8c). We observe also that the ratios $\frac{|\chi|_{max}}{|u|_{max}}$ are noticeably larger than $Ga^{-1}$, and we observe again that instabilities originating from boundary layers lead to larger interface disturbances than is predicted by the above discussed asymptotic decomposition.

The interface disturbances for all the cases considered in this section appear as short wavelength standing waves (Fig. 9). The real and imaginary parts of the interface perturbations are in counter phase for the AK20 oil – water system, and are in phase in the three other configurations. The standing waves are clearly observed in the supplied animation file. Estimations of the dimensional amplitude of the interface oscillations show that in all cases, except the AK20 oil – water system, the amplitudes remain of the order of 1 $\mu m$ or smaller (Table 6). Unexpectedly, in the AK20 oil – water system the amplitudes exceed 1 $mm$, so that this system can be a good choice for experimental study of interface waves.

It is stressed again, that in spite their very small amplitudes, including the interface perturbations in the model visibly alters the critical parameter values. In the AK20 oil – water and 3 cSt oil – FC70 systems neglecting the interface disturbances leads to qualitative changes in the patterns of most unstable disturbances of the stream function and the temperature. At the same time, if to exclude the thermocapillary force, but keep the interface deformations (Table 6), the perturbation patterns remain similar in all four cases reported in Fig. 8.

*4.3 "Small" cavity, $H = 1\ cm$, conducting horizontal boundaries*

The third and fourth series of numerical experiments were carried out for 10 times smaller cavity with $H = 1 cm$ and $L = 5 cm$. Since the Grashof and Galileo numbers are proportional to $H^3$, while the Marangoni number to $H$, reducing the cavity height increases relative effect of the thermocapillary force. The corresponding changes should be kept in mind while addressing values of the governing parameters listed in Table 1. In this section we focus on results obtained for a small cavity with perfectly conducting horizontal boundaries. The critical parameters for this case are reported in Table 4.

In a smaller cavity the critical temperature differences become larger, which happens due to decrease of the buoyancy effect (i.e., a smaller Grashof number), as well as increase of the effect of viscous friction near the boundaries.

Comparing the results obtained with and without the interface disturbances taken into account, we observe that the difference of the critical parameters is not very large, but not negligible, similarly to what was observed in the large cavity. At the same time, we observe that neglecting the thermocapillary force ($Ma = 0$) changes the critical parameters drastically. In the water – air system, neglecting the thermocapillary force makes the critical temperature difference much larger than 50°C, which brings the system out of Boussinesq model limit. It is even more striking that the ratios $\frac{|\chi|_{max}}{|u|_{max}}$ remains small, and possibly comparable with $Ga^{-1}$, for the first three systems when the full model is considered (Table 4). When the thermocapillary force is removed from the model ($Ma = 0$), the above ratios appear to be several orders of magnitude larger, contrarily from what is predicted by the asymptotic expansion (Nepomnyashchy et al., 2006). Since this observation is done for a physically inconsistent model, we do not study this issue in detail. On the other hand, these four cases present examples when in a purely buoyant convection flow the asymptotic expansion does not predict the magnitude of the interface perturbations correctly.

Estimating the dimensional amplitude of the interface disturbances (Table 6), we observe that in the small cavity the amplitudes become one to two orders of magnitude larger than in the large one for all cases except the AK20 oil – water case, for which the amplitude growth is not so large. This observation is important, since waves with the amplitudes of several tens of microns are apparently not visible, but can be detected in experiments (Kang et al., 2004; Duan et al., 2006; Zhang et al., 2014).

Comparing the flow and perturbation patterns for the large and small cavities, we observe that they are qualitatively different. The base flows and isotherms calculated at the critical temperature difference in the current case of small cavity are shown in Figs. 10 and 11. The most unstable disturbances of the stream function are plotted in Fig. 12.

In the air –water and AK20 oil – water configuration, a stronger thermocapillary force produces a strong positive horizontal velocity at the interface. As a result, action of the buoyancy force is suppressed and the counter clockwise circulation occupies the whole air layer (cf. Figs. 2a and 10a). In the four other systems, the flow patterns consisting of two clockwise circulations separated by a counter clockwise one persist, however are noticeably different from their counterparts in the large cavity. Thus, in the benzene – water and 1cSt – water systems the weaker counter clockwise circulation is located in the upper layer, and not in the lower one (cf Figs. 2b,c and 10b,c). In both cases, a large horizontal velocity at the interface completely removes secondary circulations that appeared near the vertical boundaries in the large cavity.

Apparently, differences in the velocity fields appear because of different relation of the two driving forces. The convective heat transfer driven by different velocity fields, results in qualitatively different temperature fields (cf. Figs. 3 and 11). Thus, temperature boundary layers do not develop in the first four cases (Figs. 11a,b,c,d). In the last case of the 3 cSt oil – FC70 liquid, the "hot" boundary layer produced unstable stratification near the left boundary (Fig. 11e), and the most unstable disturbance is located exactly there (Fig. 12e). In the four other cases (Fig. 12a,b,c,d), the most unstable perturbation is located in much wider regions in the bulk of both layers (Fig 12a), or only in the upper layer (Figs. 12b,c). As above, these instabilities can be possibly attributed to shear mechanisms, but it is possible also that there are additional convective mechanisms that must be studied separately. A more detailed discussion on the instability mechanisms is presented below as a separate section. Here we mention again that the instability located in the boundary layer (Fig. 12e) leads to a relatively larger interface disturbances (Table 4), as it was observed for larger cavities. A larger $\frac{|\chi|_{max}}{|u|_{max}}$ ratio observed in the AK20 oil – water system also can indicate on dominating non-buoyancy mechanisms in the instability development.

The normalized interface disturbances are reported in Fig 13. Apparently they differ from those depicted in Figs. 5 and 9. The real and imaginary parts are out of phase in all the cases. Time oscillations of the interface perturbations are visualized in the supplied animation file. In this case we observe that the interface oscillations appear as relatively long wavelength waves in the benzene – water, 1 cSt oil – water, 3 cSt oil – FC70, and AK20 oil – water systems.

In the air – water system the oscillations appear in the central part of the cavity and are characterized by a single maximum and single minimum of the elevation function.

*4.4 "Small" cavity, $H = 1\ cm$, insulating horizontal boundaries*

In the fourth series of numerical experiments we replace perfectly conducting horizontal boundaries of the small cavity by perfectly thermally insulating ones. The calculated critical parameters are summarized in Table 5. The critical temperature differences are larger than they were in the large cavity, which apparently happens for the same reasons as in the conducting boundaries case. For the benzene – water and 1 cSt oil – water system we observe a noticeable increase of critical temperature difference compared to the previous case of conducting boundaries, which is similar to our observation for large cavities.

With the absence of the thermocapillary force ($Ma = 0$), flows in the air – water and oil AK20 – water configurations remain stable with respect to the two-dimensional disturbances up to $\Delta T = 50°C$. Also in the FC30 – 3cSt oil configurations the critical temperature difference was beyond 50°C for all the cases considered. As mentioned above, further increase of the temperature difference would be meaningless, since not only applicability of the Boussinesq approximation becomes questionable, but also temperature dependence of properties of the liquids should be taken into account. It is possible, of course, that the flows are unstable with respect to the 3D perturbations, however this question is out of scope of the present study. In the other cases, the parameter $\varepsilon_1 = \beta_j(T_{max} - T_{min})$ remains of the order of $10^{-2}$, which still allows us to assume that the Boussinesq approximation remains accurate enough. Note, that excluding the interface deformations changes the critical temperature differences within ≈ 10% range, similarly to what was observed above. At the same time excluding the thermocapillary force from the mode leads to drastic changes in $\Delta T_{cr}$, showing the main driving in this model is due to the Marangoni force.

Estimations of the dimensional amplitude of the interface oscillations (Table 6) show that contrarily to the previous case, the amplitudes in the small cavity do not increase so drastically, compared to change of the boundary conditions of the large cavities. At the same time, we do observe a drastic increase of the amplitudes in the model with excluded thermocapillary force. Thus, we observe here a quite unexpected suppression of interface oscillations by the thermocapillary forcing.

Comparing the flow patterns (Fig. 14) in this and the previous case, we observe that the patterns of the air – water flow remain similar, while in the other systems the counter clockwise circulations become weaker. This can be explained either by decrease of the thermocapillary force along the interface, or by increase of buoyancy forces in the upper layer. Comparing the isotherm patterns (cf. Fig. 15 and Fig. 11), we observe steeper changes of the temperature in the horizontal direction for the benzene – water and 1 cSt oil – water systems. In the AK20 oil – water system (Fig. 15d) change of the temperature along the interface is less steep that in the conducting boundaries case (Fig. 11d). In the air – water system, the isotherm patterns differ strongly (cf. Fig. 11a and Fig. 15a), however, the total balance of forces produces similar flow patterns (cf. Fig. 10a and Fig. 14a).

Patterns of the most unstable disturbances of the stream functions (Fig. 16) show that in the three systems, benzene – water, 1 cSt oil – water, and AK20 oil – water (Fig. 16b,c,d), the instability develops several vortices in the upper layer. This is similar to what was observed in these systems in the small cavity with conducting boundaries (Fig.12). In the large cavity with conducting boundaries we observed development of similar vortical structures in the lower layer for air – water and 1 cSt oil – water systems (Fig. 4). In the air – water system the instability develops almost symmetrically in both layers also similarly to the previous case of conducting horizontal boundaries. We have attributed these instabilities to shear mechanisms, and will present a more detail discussion in the following section.

The normalized interface disturbances are reported in Fig 17 and in the supplied animation file. They are apparently different from those depicted in Figs. 5, 9 and 13. The real and imaginary parts are out of phase in all the cases. Only 1 cSt – water system produces interface waves similar to those observed for the small cavity with conducting horizontal boundaries. In the case of air – water and AK20 oil –water systems the single minimum and maximum oscillations spread over the whole interface from the left to the right vertical boundary. In the benzene – water system these oscillations are modulated by ones with a shorter wavelength.

*4.5 More insight into most unstable disturbances*

All the perturbation patterns reported in Figs. 4, 8, 12, and 16 can be divided into three groups. In the first group the most unstable perturbations are located near one of the vertical boundaries (Figs. 4a,b,d,e 8a-8d, 12e), where the instability is attributed to the boundary layers and/or to the local unstable stratification discussed above. Such localized disturbance was reported for a single-phase flow by Gelfgat (2007), where difficulties in its numerical resolution were discussed. The present results show also that instabilities of this kind create interface deformations with the largest amplitudes observed.

In the second group, the most unstable disturbances appear almost symmetrically in both layers. These patterns are observed only in the small cavity with $H = 1\ cm$ and only for the air –water system (Figs. 12a and 16a). To speculate about a possible origin of instability in these cases we plot in Fig. 18 vertical profiles of the horizontal velocity across horizontal location of maximum of the perturbation absolute value. We observe that in both layers the velocity profile resembles a smooth return flow similar to the classical Birikh (1966) profile. Since interface deformations do not strongly affect the critical numbers (Tables 4 and 5), we expect that instability of these flows develops similarly to their single layer counterparts. These instabilities are discussed, e.g., in Davis (1987), Gershuni et al. (1989), and Priede & Gerbeth (1997). In the infinite layers these instabilities appear as three-dimensional oblique waves (Davis, 1987). Within the present two-dimensional model the possible 3D disturbances were not studied. The 2D disturbances in the discussed cases appear as smaller scale vortical structures, which is similar to 2D perturbations in the infinite layer single phase flow, where such perturbations were called "two-dimensional waves" (Davis, 1987), and are usually reported as "transverse rolls". This mode of instability was visualized experimentally by Riley & Neitzel (1998).

The third group of the dominant perturbations appear as a wavy vortical structure in only one of the layers, i.e., in the lower layer in 1 cSt oil – water system in large cavity with conducting boundaries (Fig. 4c), and in upper layer in the benzene – water,1 cSt oil – water, AK20 oil – water systems in small cavities (Figs. 12b,c,d and 16b,c,d). To speculate about possible reason fort these instabilities, we plot the corresponding profiles of the base flow $u(z)$ and absolute value of the most unstable disturbance $\tilde{u}(z)$ (Fig. 19). The profiles correspond to the *x*-location of the maximum of the absolute value of the corresponding perturbation (Figs. 4, 12, and 16). The common property of the base flow profiles (Fig. 19a) is a presence of the inflection point, where $\partial^2 u/\partial z^2 = 0$, on each of them. The presence of the inflection point together with the change of flow direction resembles configuration of mixing layer, which is subject to the Kelvin-Helmholtz instability mechanism (Chadrasekhar, 1961). In non-isothermal (density stratified) mixing layers the classical Kelvin-Helmholtz instability is replaced by the Holmboe instability (Holmboe, 1962). Gelfgat & Kit (2006) argued that both instabilities results from the same shear mechanism, where the Kelvin-Helmholtz modes are characterized by two, stable and unstable, real eigenvalues, which, with the increasing stratification, may coincide forming a complex conjugate pair. In this case the steady Kelvin-Helmholtz disturbance turns into waves traveling in opposite directions and known as Holmboe instability.

To illustrate the following argument, we show an example of the $u = tanh(z/\delta)$ velocity profile and a corresponding Holmboe mode of the stratified mixing layer in an insert of Fig. 19b. The data is taken form Gelfgat & Kit (2006). The absolute value of the Holmboe modes velocity profile has a sharp maximum shifted from the inflection point location ($x = 0$) to smaller or larger $x$. These shifts make the Holmboe modes different from the Kelvin-Helmholtz modes, whose sharp maxima are located exactly at the inflection point, while the whole shape of the disturbances absolute value profiles remain similar for the Kelvin-Helmholtz and Holmboe modes, having two more local maxima at both sides of the main sharp maximum (Gelfgat & Kit, 2006). The latter local maxima are symmetric in the Kelvin-Helmholtz temporal modes, and are slightly asymmetric in the Holmboe modes.

Observing the profiles of absolute value of the most unstable disturbance $\tilde{u}(z)$ (Fig. 19b) we notice that the profiles of the benzene – water and 1 cSt oil – water systems replicate the features of the classical Holmboe mode shown in the insert. Namely, they have a main maximum shifted from the inflection point of the corresponding velocity profile (Fig. 19a), and two smaller maximal located at both sides of the main one. The AK20 oil – water system profiles have a more complicated shape of the main maximum, but also have two smaller local maxima at both sides of the main one. This allows us to assume that the observed instabilities develop in the same way as Holmboe modes in stratified mixing layers. Obviously, this observation is not sufficient to make a definite conclusion.

## 5. Conclusions and discussion

The main objective of this study was examination of the effect of interface disturbances on stability of buoyant/thermocapillary convective flows. To address this objective, we have studied a relatively simple configuration, restricted to the 2D model, but taking into account properties of liquids used in previously published experiments. We have found that in some systems effect of the interface disturbances is negligible, while in others it can alter the critical temperature difference by approximately 10% (Tables 2-5). Theoretically, this effect can be larger in other systems, so that to arrive to a correct quantitative answer, the interface disturbances should be taken into account.

At the same time, we did not find a case where including the interface disturbances into the model led to a qualitative change of the pattern of the most unstable perturbation of the stream function or temperature. This means that in the configurations considered the interface disturbances were not the reason triggering instability, but only affected other destabilizing mechanisms. It is worth to mention that in many cases reported in Tables 2-5, the interface disturbances exhibited a quite unexpected stabilizing effect, so that the corresponding critical temperature differences slightly grew compared to the non-deformable interface model. This stabilization can result from an additional energy needed to deform a capillary interface.

Basing on the obtained stability results we tried to estimate the dimensional amplitude of the interface oscillations in slightly supercritical flow regimes. To do this we assumed that the amplitude of the horizontal velocity supercritical oscillations reaches 1% of the maximum value of the steady slightly subcritical horizontal velocity. The latter yields the dimensionless amplitude of the whole perturbation that includes velocity, temperature, and pressure fields, as well as interface deformations. Multiplying the interface disturbance by the estimated amplitude, and rescaling the result to the dimensional value, we arrived to the estimation of the amplitude (Table 6). We have found that in some systems the amplitudes are estimated to be below 1 $\mu m$, while in the others they can reach tens of microns. In the latter case the interface oscillations become experimentally detectable, which allows for non-intrusive measurements of the instability onset (Kang et al., 2004; Duan et al., 2006; Zhang et al., 2014, Nezhihovski et al., 2022). The interface disturbance can appear as standing or travelling waves, whose wavelength varies from a very short (e.g., Fig 9) to the quite large restricted only by length of the cavity (e.g., Fig. 17). It is quite unexpected, that the estimated amplitudes do not directly correlate with the difference in the critical values in the models with and without interface disturbances.

It should be noted that the largest amplitudes of the interface deformations were observed for the AK20 oil – water system. Thus, in a large cavity with thermally insulating boundaries, the amplitude is estimated to exceed 1 $mm$. One of the obvious reasons for that is close density values, 950 and 1000 $kg/m^3$, of the two liquids (Table 1). On the other hand, these values may be not close enough to produce such striking effect, and there are additional reasons that yet should be studied.

The asymptotic expansion of the two-phase flow in a power series of $\varepsilon_1 = \beta_j(T_{max} - T_{min})$, usually used to derive the Boussinesq model, leads to the conclusion that the interface deformation amplitudes are estimated as $\varepsilon_1/Ga$, where $Ga \gg 1$. Basing on that, we expected to observe a very small amplitude of the interface disturbances with a possible effect of the latter on the resulting instability, as follows from Eq. (14) and the corresponding discussion.

However, our observation show that this estimation can be applied only when instability is driven mainly by the buoyancy forces. In other cases, when instability originates from boundary layers or from velocity shear in the bulk of the flow, the interface disturbances can become significantly larger. Altogether, the present observations show that for a combined buoyancy/thermocapillary double layer connective flow, a complete linear stability problem should include disturbances of the interface.

It is noteworthy that all the three main patterns sketched in Fig. 1 were observed in different two-layer systems. Clearly, since the flow patterns are qualitatively different, one would expect different routes to their primary instability. It should be stressed here that a complete answer on the flow stability properties can be obtained only considering full three-dimensional model with fully 3D base flow and all possible 3D disturbances. Such studies are becoming feasible for single phase convective flows (see, e.g., Gelfgat, 2020) and can be extended to the two-phase ones, however, they are still too CPU time consuming to perform a series of numerical experiments. In our opinion, a study of such kind can be meaningful if is carried out in connection with a reliable experiment, for which all the thermophysical and geometrical parameters are defined. This is beyond the scope of present study.

In the framework of this study we do observe qualitatively different most unstable two-dimensional disturbances, which can be divided into three groups. In the first group the perturbations are localized near one of the boundaries, were we observe two potentially destabilizing features: thin boundary layers and unstable temperature stratification. The second group exhibits Birikh-like smooth profiles in each layer, so that the instability sets in almost symmetrically in both layers due to the transverse rolls disturbances, which are characteristic for single layer flows of this kind. In the third group the instability develops only in one of the layers and can be described as vortical structures travelling along the interface. The base flow vertical velocity profiles in these cases have inflection points, which indicates on a possible inviscid instability mechanism. We presented some arguments showing that these instabilities exhibit certain similarities with Holmboe instability observed in stratified mixing layers.

**Acknowledgment**
This research was supported by Israel Science Foundation (ISF) grant No 415/18.
The author is thankful to A. Oron and A. Nepomnyashchy for important and fruitful discussions.

**References**

Afrid, M., and Zebib, A. 1990 Oscillatory three-dimensional convection in rectangular cavities and enclosures. Phys. Fluids A, 2, 1318-1327.

Ben Hadid, H., and Roux, B. 1992 Buoyancy- and thermocapillary-driven flows in differentially heated cavities for low-Prandtl-number fluids. J. Fluid. Mech., 235, 1-36.

Birikh, R. V. 1966 Thermocapillary convection in a horizontal layer of liquid. J. Appl. Mech. Tech. Phys., 3, 69-72.

Birikh, R.V. and Bushueva, S.V. 2001 Thermocapillary Instability in a Two-Layer System with a Deformable Interface. Fluid Dynamics, 36, 349-355.

Burguete, J., Mukolobwiez, N., Davidaud, F., Garnier, N., and Chiffaudel, A. 2001 Buoyant-thermocapillary instabilities in extended liquid layers subjected to a horizontal temperature gradient. Phys. Fluids, 13, 2773-2787.

Castillo, J.L. and Velarde, M. G. 1982 Buoyancy-thermocapillary instability: the role of interfacial deformation in one- and two-component fluid layers heated from below or above. J. Fluid. Mech., 125, 463-464.

Cerisier, P., Jamond, C., Pantaloni, J., and Charmet, J.C. 1984 Déformation de la surface libre en convection de Bénard–Marangoni. J. Physique, 45, 405-411.

Chandrasekhar S. 1961. Hydrodynamic and Hydromagnetic Stability. Oxford, 652 pp.

Chen, J.-C. and Hwu, F.-S. 1993 Oscillatory thermocapillary flow in a rectangular cavity. Int. J. Heat Mass Transfer, 36, 3743-3749.

Davis, S.H., 1987 Thermocapillary instabilities. Ann. Rev Fluid Mech., 19, 403-435.

Duan, L., Kang, Q., and Hu, W. R. 2006. Characters of surface deformation and surface wave in thermal capillary convection. Science in China, Series E: Technological Sciences, 49, 601-610.

Fedyushkin A. I. 2020 The effect of convection on the position of the free liquid surface under zero and terrestrial gravity. J. Phys. Conference Series, **1675**, 012039.

Gelfgat, A., Bar-Yoseph, P. Z. and Yarin, A. L. 1997 On oscillatory instability of convective flows at low Prandtl number, J Fluids Engineering 119, 823-830.

Gelfgat A., Kit E. 2006 Spatial versus temporal instabilities in parametrically forced stratified mixing layer. J. Fluid. Mech., **552**, 189-227.

Gelfgat, A. 2007 Three-dimensional instability of axisymmetric flows, solution of benchmark problems by a low-order finite volume method. Int. J. Numer. Meths. Fluids 54, 269-294.

Gelfgat, A. 2020 Instability of natural convection in a laterally heated cube with perfectly conducting horizontal boundaries. Theor. Comput. Fluid Dyn. 34, 693-711.

Gelfgat, A., Brauner, N., 2020. Instability of stratified two-phase flows in rectangular ducts. Int. J. Multiphase Flow 131, 103395.

Gershuni ,G.Z. and Zhukhovitskii ,E.M. 1976 Convective Stability of Incompressible Fluids, Keter Publishing House, Jerusalem, 330 pp.

Gershuni, G.Z., Zhukhovitskii, E.M., and Nepomnyashchy, A. 1989 Stability of Convective Flows. Moscow, Nauka, 320 pp.

Golovin, A. A, Nepomnyashchy, A. A., and Pismen, L. M. 1995. Pattern formation in large-scale Marangoni convection deformable interface. Physica D, 81, 117-147.

Hamed, M. and Floryan, J. M. 2000 Marangoni convection. Part 1. A cavity with differentially heated sidewalls. J. Fluid Mech., 405, 79-110.

Holmboe, J. 1962 On the behaviour of symmetric waves in stratified shear flows. Geofys. Publ. 24, 67-113.

Huang J.-J., Huang H., Wang X. 2014 Numerical study of drop motion on a surface with stepwise wettability gradient and contact angle hysteresis. Phys. Fluids, **26**, 062101.

Kang, Q., Duan, L., and Hu, W. R. 2004 Experimental study of surface deformation and flow pattern on buoyant-thermocapillary convection. Microgravity Sci. Technol., 15/2, 18-24.

Kuhlmann, H. and Albensoeder, S. 2008 Three-dimensional flow instabilities in a thermocapillary-driven cavity. Phys. Rev. E, 77, 036303.

Landau L. D., Lifshitz E.M. 1987 Fluid Mechanics, Oxford, 551 pp.

Lappa, M. 2009. Thermal convection: Patterns, Evolution and Stability. John Wiley & Sons, Sussex, UK, 670 pp.

Laure, P. and Roux, B. 1989 linear and non-linear analysis of the hadley circulation, J. Cryst. Growth, 97, 226-234.

Lebon, G., Dauby, P C. and Regnie V.C. 2001 role of interface deformations in benard-marangoni instability. Acta Astronautica, 48, 617-627.

Li, Y., Grogoriev, R. and Yoda, M. 2014 Experimental study of the effect of noncondensables on buoyancy-thermocapillary convection in a volatile low-viscosity silicone oil. Phys. Fluids, 26, 122112.

Liu, Q.S., Chen, G., and Roux, B. 1993 Thermogravitational and thermocapillary convection in a cavity containing two superposed immiscible liquid layers. Int. J. Heat Mass Transfer, **36**, 101-117.

Liu, Q.-S., Zhou, B.-H., Liu, R., Nguen-Thi, H., and Billia B. 2006 Oscillatory instabilities of two-layer Rayleigh-Marangoni-Bénard convection. Acta Astronautica, 59, 40-55.

Lyubimov D. V., Lyubimova T. P., Alexander J. I. D., Lobov N. I. 1998 On the Boussinesq approximation for fluid systems with deformable surfaces. Adv. Space Res., **22**, 119-1168.

Lyubimova T. P., D.V. Lyubimov D. V., and Parshakova, Y.N. 2015 Implications of the Marangoni effect on the onset of Rayleigh–Benard convection in a two-layer system with a deformable interface. Eur. Phys. J. Special Topics, 224, 249-259.

Madruga, S., Pérez-Garcia, C., and Lebon, G 2003 Convective instabilities in two superposed horizontal liquid layers heated laterally. Phys. Rev. E, 68, 041607.

McLelland, M. A. 1995 time-dependent liquid metal flows with free convection and a deformable free surface. Int. J Numer. Meths. Fluids, 20, 603-620.

Mihaljan, J. M 1962 A rigorous exposition of the Boussinesq approximations applicable to a thin layer of fluid. Astrophys. J., 136, 1126-1133.

Mundrane, M. and Zebib, A. 1994 Oscillatory buoyant thermocapillary flow, Phys. Fluids, 6, 3294-3305.

Nepomnyashchy A., Simanovskii I., Legros J. C. 2006 Interfacial Convection in Multilayer Systems. Springer, NY, 320 pp.

Nepomnyashchy A. A. and Simanovskii I. B. 2006 Nonlinear development of oscillatory instability in a two-layer system under the combined action of buoyancy and thermocapillary effect, J. Fluid Mech., 555, 177-202.

Mundrane, M., Xu, J., and Zebib, A. 1995 thermocapillary convection in a rectangular cavity with a deformable interface. Adv. Space Res., 16, 41-53.

Nejati, I., Dietzel, M., and Hardt, S. 2015 Conjugated liquid layers driven by the short-wavelength Bénard-Marangoni instability: experiment and numerical simulation. J. Fluid Mech., 783, 46-71.

Nezihovski, Y., Gelfgat, A., Ullmann, A., and Brauner N. 2022 Experimental measurements versus linear stability analysis for primary instability of stratified two-phase flows in a square rectangular duct. Int. J. Multiphase Flows, to appear, https://doi.org/10.1016/j.ijmultiphaseflow.2022.104061

Pamentier, P. M., Regnier, V. C., and Lebon, G. 1993 Buoyant-thermocapillary instabilities in medium-Prandtl-number fluid layers subject to a horizontal temperature gradient. Int. J. Heat Mass Transfer, 36, 2417-2427.

Patne, R., Agnon, Y., and Oron, A. 2020 Thermocapillary instabilities in a liquid layer subjected to an oblique temperature gradient. J. Fluid Mech., 906, A12-1 – A12-30.

Pérez-Garcia, C. and Carneiro, G. 1990 Linear stability analysis of Bénard-Marangoni convection in fluids with a deformable free surface. Phys. Fluids A, 3, 292-298.

Priede, J. and Gerbeth, G. 1997 Convective, absolute, and global instabilities of thermocapillary-buoyancy convection in extended layers. Phys. Rev. E, 56, 4187-4199.

Rasenat, S., Busse, FH., and Rehberg, I. 1989 A theoretical and experimental study of double-layer convection. J. Fluid Mech., 199, 519-540.

Regnier, V.C., Dauby, P. C., and Lebon, G. 2000 Linear and nonlinear Rayleigh–Bénard–Marangoni instability with surface deformations. Phys. Fluids, 12, 2787-2799.

Renardy, M. and Renardy, Y. 1988 Bifurcating solutions at the onset of convection in Bénard problem for two fluids. Physica D, 32, 227-252.

Riley, R. J. and Neitzel, G. P. 1998 Instability of thermocapillary-buoyancy convection in shallow layers. Part 1. Characterization of steady and oscillatory instabilities. J. Fluid Mech., 359, 143-164.

Sáenz, P. J., Valluri, P., Sefiane, K., Karapetsas, G., and Matar O. K. 2013 Linear and nonlinear stability of hydrothermal waves in planar liquid layers driven by thermocapillarity. Phys. Fluids, 25, 094101.

Schatz, M F. and Neitzel, G. P. 2001 Experiments on thermocapillary instabilities. Annu. Rev. Fluid Mech., 33, 93-127.

Scriven, LE. And Sternling, C.V. 1964 On cellular convection driven by surface-tension gradients: effects of mean surface tension and surface viscosity. J. Fluid Mech., 19, 321-340.

Simanovskii, I. B., Viviani, A., Dubois, F, and Legros, J.-C. 2012 The influence of the horizontal component of the temperature gradient on nonlinear convective oscillations in two-layer systems. Phys. Fluids, 24, 102-108.

Smith, M.K. and Davis, S.H. 1983b Instabilities of dynamic thermocapillary liquid layers. Part 2. Surface-wave instabilities. J. Fluid Mech., 132, 145-162.

Smith, M.K. and Davis, S.H. 1983a Instabilities of dynamic thermocapillary liquid layers. Part 1. Convective instabilities. J. Fluid Mech., 132, 119-144.

Straub J., Weinzierl A., Zell M. 1990 Thermokapillare Grenzflächenkonvektion an in einem Temperaturgradientenfeld. Wärme- und Stoffübertragung, **25**, 281-288.

Vanhaelen, Q. 2019 Thermo-capillary effects along a deformable singular interface between two immiscible fluids. Physica A, 531, 121803.

Velarde M.G., Nepomnyashchy A. A., Hennenberg M. 2001 Onset of oscillatory interfacial instability and wave motions in Bénard layers. Adv. Appl. Mech., 37, 167-238.

Vila, M. A., Kuz, V. A., Garazo, A. N., and Rodriguez, A. E. 1987 Marangoni instability, effects of tangential surface viscosity on a deformable interface. J. Physique, 48, 1895-1900.

Villers, D. and Platten, J.K. 1990 Influence of interfacial tension gradients on thermal convection in two superposed immiscible liquid layers. Appl. Sci. Res., 47, 177-191.

Wahal, S. and Bose, A. 1988 Rayleigh-Bénard and interfacial instabilities in two immiscible liquid layers Phys. Fluids, 31, 3502-3510.

Wang, P. and Kawahita, R. 1996 Transient buoyancy-thermocapillary convection in two superposed immiscible liquid layers. Numerical Heat Transfer, Pt. A, 30, 477-501.

Wang, P. and Kawahita, R. 1998 Oscillatory behavious on buoyancy-thermocapillary convection in fluid layers with a free surface. Int J. Heat Mass Transfer, 41, 399-409.

Zeren R. W., Reynolds W.C. 1972 Thermal instabilities in two-fluid horizontal layers. J. Fluid. Mech., **305**, 305-327.

Zeytonian, R. Kh. 2003 Joseph Boussinesq and his approximation: a contemporary view. C. R. Mechanique, 331, 575-586.

Zhang, L, Duan, L, and Kang, Q. 2014 An experimental research on surface oscillation of buoyant-thermocapillary convection in an open cylindrical annuli. Acta Mechanica Sinica, 30, 681-683.

Zhao A. X, Wagner C, Narayanan R., Friedrich R 1995 Bilayer Rayleigh-Marangoni convection: transitions in flow structures at the interface. Proc. R. Soc. Lond. A, **451**, 487-502.

Table 1. Properties of liquids used in calculations. The dimensionless parameters are calculated for $H = 10\ cm$

| Liquid | $\rho\left(\frac{kg}{m^3}\right)$ | $\mu\left(\frac{kg}{m \cdot s}\right) \cdot 10^3$ | $k\left(\frac{W}{m \cdot K}\right)$ | $c_p\left(\frac{J}{kg \cdot K}\right)$ | $\beta\left(\frac{1}{K}\right) \cdot 10^4$ | $\sigma\left(\frac{N}{m}\right)$ with water | $\gamma\left(\frac{N}{m \cdot K}\right) \cdot 10^4$ with water | $Gr/\Delta T \cdot 10^{-6}$ | $Mn/\Delta T \cdot 10^{-3}$ | $Ga \cdot 10^{-8}$ | $Bo$ |
|---|---|---|---|---|---|---|---|---|---|---|---|
| Water | 1000 | 1.0 | 0.6019 | 4181 | 2.27 | | | 2.23 | | 98.1 | |
| Air | 1 | 0.018 | 0.026 | 1006 | 34.3 | 0.072[1] | 1.0[1] | 0.0104 | 3.086 | 0.303 | 1.3625 |
| Benzene | 879 | 0.649 | 0.144 | 1729 | 1.25 | 0.03418[1] | 0.56[1] | 2.25 | 11.687 | 180.0 | 2522.8 |
| 1 cSt oil | 820 | 0.82 | 0.1 | 2000 | 1.29 | 0.0169[1] | 0.6[1] | 1.27 | 7.32 | 98.1 | 4759.9 |
| Oil AK20 | 950 | 19.0 | 0.14 | 1440 | 9.7 | 0.03282[1] | 1.22[1] | 0.00238 | 0.0321 | 0.245 | 2839.6 |
| 3 cSt oil | 925 | 9.27 | 0.125 | 1100 | 11.0 | | | | | | |
| FC-70 | 1940 | 27.16 | 0.07 | 1555 | 10.0 | 0.005[2] | 0.32[2] | 0.00127 | 0.0732 | 0.981 | 4759.9 |

[1] – with water;   [2] – with 3 cSt oil

Table 2. $A = \frac{L}{H} = 5, H = 10 cm,$ grid $2000 \times 301$. Same heating from the side, conducting horizontal boundaries.

| Liquids | $\Delta T_{cr}$ | $\omega_{cr}$ | $|\chi|_{max}/|u|_{max}$ | $\Delta T_{cr}(\chi = 0)$ | $\omega_{cr}(\chi = 0)$ | $\Delta T_{cr}(Ma = 0)$ | $\omega_{cr}(Ma = 0)$ | $\frac{|\chi|_{max}}{|u|_{max}}, Ma = 0$ |
|---|---|---|---|---|---|---|---|---|
| Air - water | 0.1676 | 0.5843 | 7.4·10$^{-06}$ | 0.1677 | 0.5843 | 0.1835 | 0.5506 | 1.3·10$^{-05}$ |
| Benzene - water | 0.01539 | 1.5837 | 2.0·10$^{-05}$ | 0.01542 | 1.5778 | 0.01541 | 1.5821 | 2.1·10$^{-05}$ |
| 1 cSt oil – water | 0.003321 | 0.002399 | 3.7·10$^{-06}$ | 0.003806 | 0.02994 | 0.0033603 | 0.02519 | 3.9·10$^{-06}$ |
| Oil AK20 – water | 0.1385 | 0.2929 | 0.0281 | 0.1587 | 0.2981 | 0.1362 | 0.2926 | 0.0286 |
| 3 cSt oil – FC70 | 0.04726 | 0.07695 | 2.4·10$^{-04}$ | 0.04725 | 0.07692 | 0.04767 | 0.07845 | 2.4·10$^{-04}$ |

Table 3. $A = \frac{L}{H} = 5, H = 10 cm,$ grid $2000 \times 301$. Same heating from the side, insulated horizontal boundaries.

| Liquids | $\Delta T_{cr}$ | $\omega_{cr}$ | $|\chi|_{max}/|u|_{max}$ | $\Delta T_{cr}(\chi = 0)$ | $\omega_{cr}(\chi = 0)$ | $\Delta T_{cr}(Ma = 0)$ | $\omega_{cr}(Ma = 0)$ | $\frac{|\chi|_{max}}{|u|_{max}}, Ma = 0$ |
|---|---|---|---|---|---|---|---|---|
| Air - water | >50 | | >50 | | | >50 | | |
| Benzene - water | 0.05687 | 1.2591 | 1.3·10$^{-04}$ | 0.05550 | 1.1999 | 0.06543 | 1.2647 | 1.4·10$^{-04}$ |
| 1 cSt oil – water | 0.09137 | 0.8407 | 9.0·10$^{-05}$ | 0.08952 | 0.8707 | 0.03767 | 0.9699 | 4.7·10$^{-05}$ |
| Oil AK20 – water | 1.8456 | 0.2646 | 0.136 | 1.5676 | 0.2464 | 2.7267 | 0.3125 | 0.233 |
| 3 cSt oil – FC70 | 1.6356 | 0.08192 | 0.00412 | 1.7577 | 0.08258 | 1.7085 | 0.08328 | 0.00456 |

Table 4. $A = \frac{L}{H} = 5, H = 1cm,$ grid $2000 \times 301$. Same heating from the side, conducting horizontal boundaries.

| Liquids | $\Delta T_{cr}$ | $\omega_{cr}$ | $|\chi|_{max}/|u|_{max}$ | $\Delta T_{cr}(\chi = 0)$ | $\omega_{cr}(\chi = 0)$ | $\Delta T_{cr}(Ma = 0)$ | $\omega_{cr}(Ma = 0)$ | $\frac{|\chi|_{max}}{|u|_{max}}, Ma = 0$ |
|---|---|---|---|---|---|---|---|---|
| Air - water | 3.5284 | 0.6904 | 1.2·10⁻⁰⁶ | 3.5400 | 0.6895 | >50 | | |
| Benzene - water | 2.0488 | 0.5565 | 1.1·10⁻⁰⁴ | 1.8375 | 0.4695 | 2.9581 | 0.02755 | 0.0065 |
| 1 cSt oil – water | 1.0319 | 0.2445 | 2.7·10⁻⁰⁴ | 0.9304 | 0.2031 | 3.3688 | 0.02534 | 0.0235 |
| Oil AK20 – water | 12.5566 | 0.06644 | 0.011 | 11.4453 | 0.05128 | 40.2748 | 0.007440 | 0.624 |
| 3 cSt oil – FC70 | 36.8585 | 0.07742 | 0.0576 | 36.6357 | 0.07773 | 44.9267 | 0.05400 | 0.278 |

Table 5. $A = \frac{L}{H} = 5, H = 1cm,$ grid $2000 \times 301$. Same heating from the side, insulated horizontal boundaries.

| Liquids | $\Delta T_{cr}$ | $\omega_{cr}$ | $|\chi|_{max}/|u|_{max}$ | $\Delta T_{cr}(\chi = 0)$ | $\omega_{cr}(\chi = 0)$ | $\Delta T_{cr}(Ma = 0)$ | $\omega_{cr}(Ma = 0)$ | $\frac{|\chi|_{max}}{|u|_{max}}, Ma = 0$ |
|---|---|---|---|---|---|---|---|---|
| Air - water | 3.7894 | 0.7432 | 2.0·10⁻⁰⁶ | 3.8652 | 0.7379 | >50 | | |
| Benzene - water | 7.07494 | 0.5505 | 0.0011 | 6.6036 | 0.046598 | 0.6411 | 0.2003 | 0.0011 |
| 1 cSt oil – water | 2.8457 | 0.4136 | 1.45·10⁻⁰⁴ | 2.5585 | 0.4493 | 38.2705 | 0.9709 | 1.45·10⁻⁰⁴ |
| Oil AK20 – water | 12.5496 | 0.06392 | 2.94·10⁻⁰⁵ | 11.4381 | 0.05124 | >50 | | |
| 3 cSt oil – FC70 | >50 | | | >50 | | >50 | | |

Table 6. Estimation of interface oscillations amplitude (in $\mu m$) assuming amplitude of oscillations of the horizontal velocity to be 0.01 of the horizontal velocity maximal value.

| System | $H = 10\ cm$ | | $H = 1\ cm$ | | $H = 10\ cm, Mn = 0$ | | $H = 1\ cm, Mn = 0$ | |
|---|---|---|---|---|---|---|---|---|
| | Conducting boundaries | Insulating boundaries | Conducting boundaries | Insulating boundaries | Conducting boundaries | Insulating boundaries | Conducting boundaries | Insulating boundaries |
| Air – water | 0.023 | | 0.185 | 0.132 | | | | |
| Benzene – water | 0.597 | 0.351 | 23.9 | 0.247 | 0.438 | 0.426 | 56.6 | 9.0 |
| Oil 1 cSt – water | 0.0064 | 0.243 | 0.693 | 0.262 | 0.028 | 0.079 | 277.0 | 11.5 |
| Oil AK20 – water | 30.5 | 1780 | 40.3 | 70.8 | 676.0 | 1420 | 6950 | |
| FC70 – Oil 3 cSt | 0.15 | 0.250 | 47.3 | | 0.148 | 4.19 | 159.0 | |

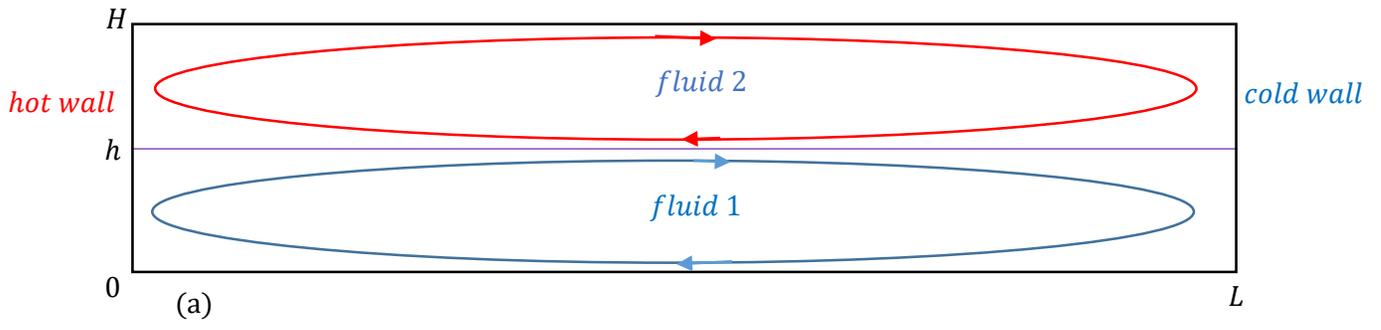

(a)

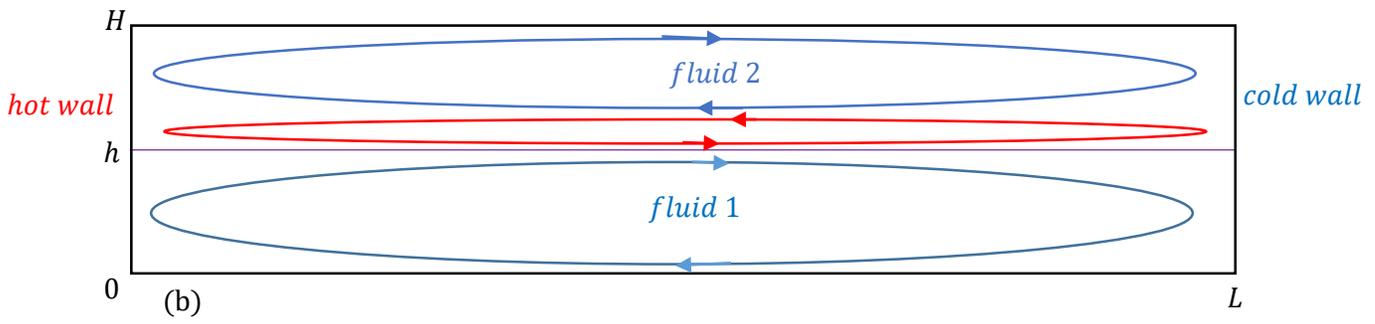

(b)

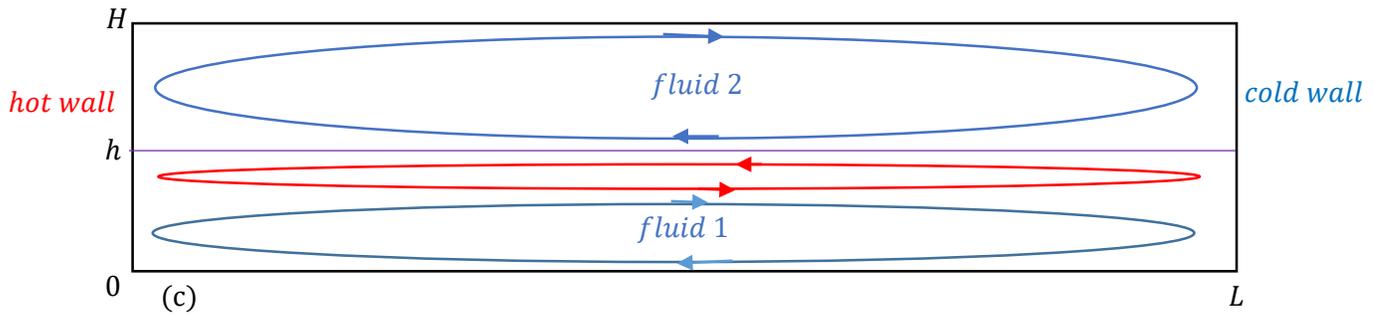

(c)

Fig. 1. Sketch of the geometry and the flow. Clockwise and counter clockwise circulations are shown in blue and red color, respectively.

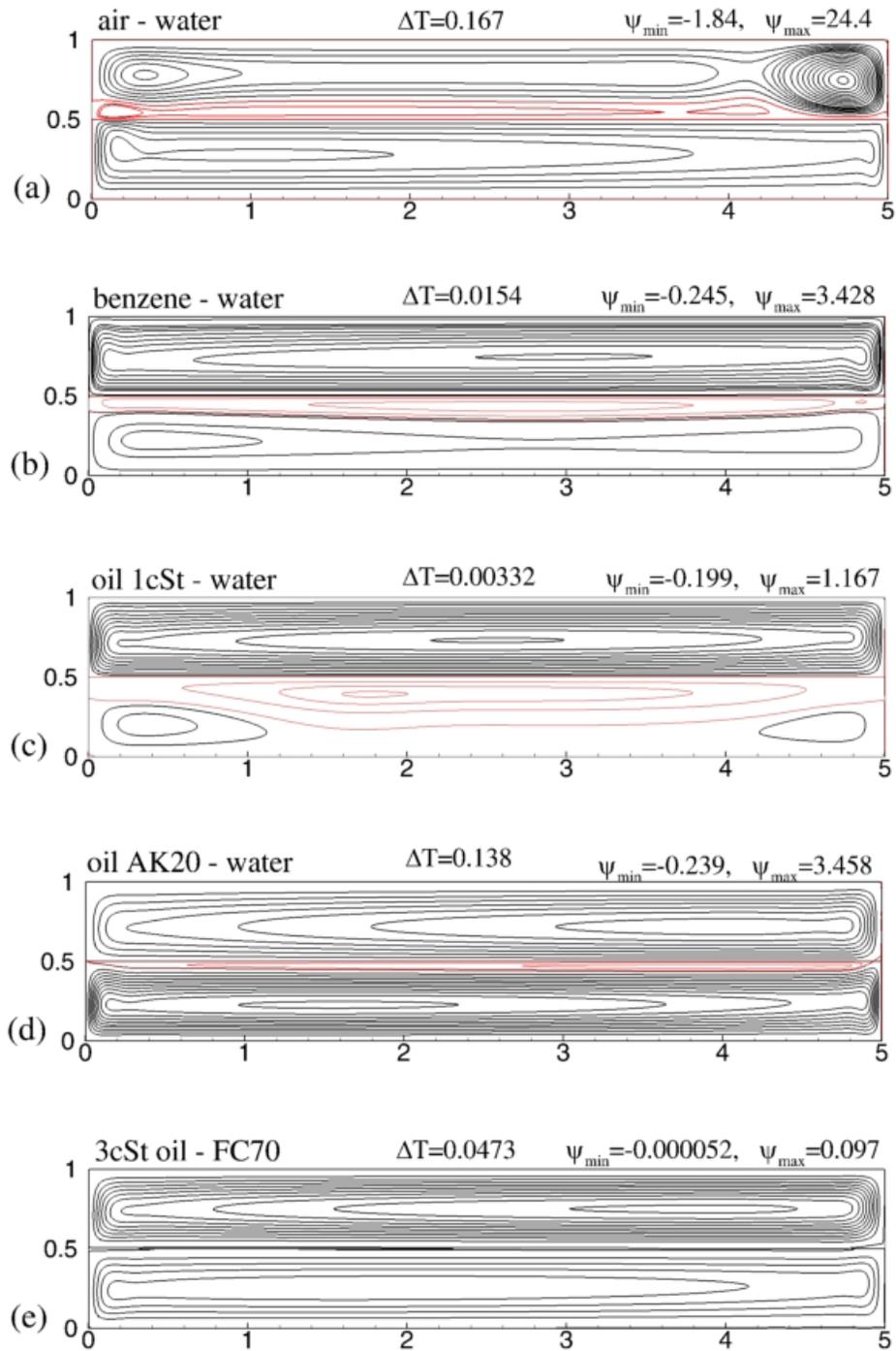

Fig. 2. Streamlines of the considered two fluid systems near their critical points. $H = 10 cm$, perfectly conducting horizontal boundaries. Uniform temperatures at the vertical boundaries. The black streamlines show clockwise circulations, while the red streamlines show counter clockwise circulation.

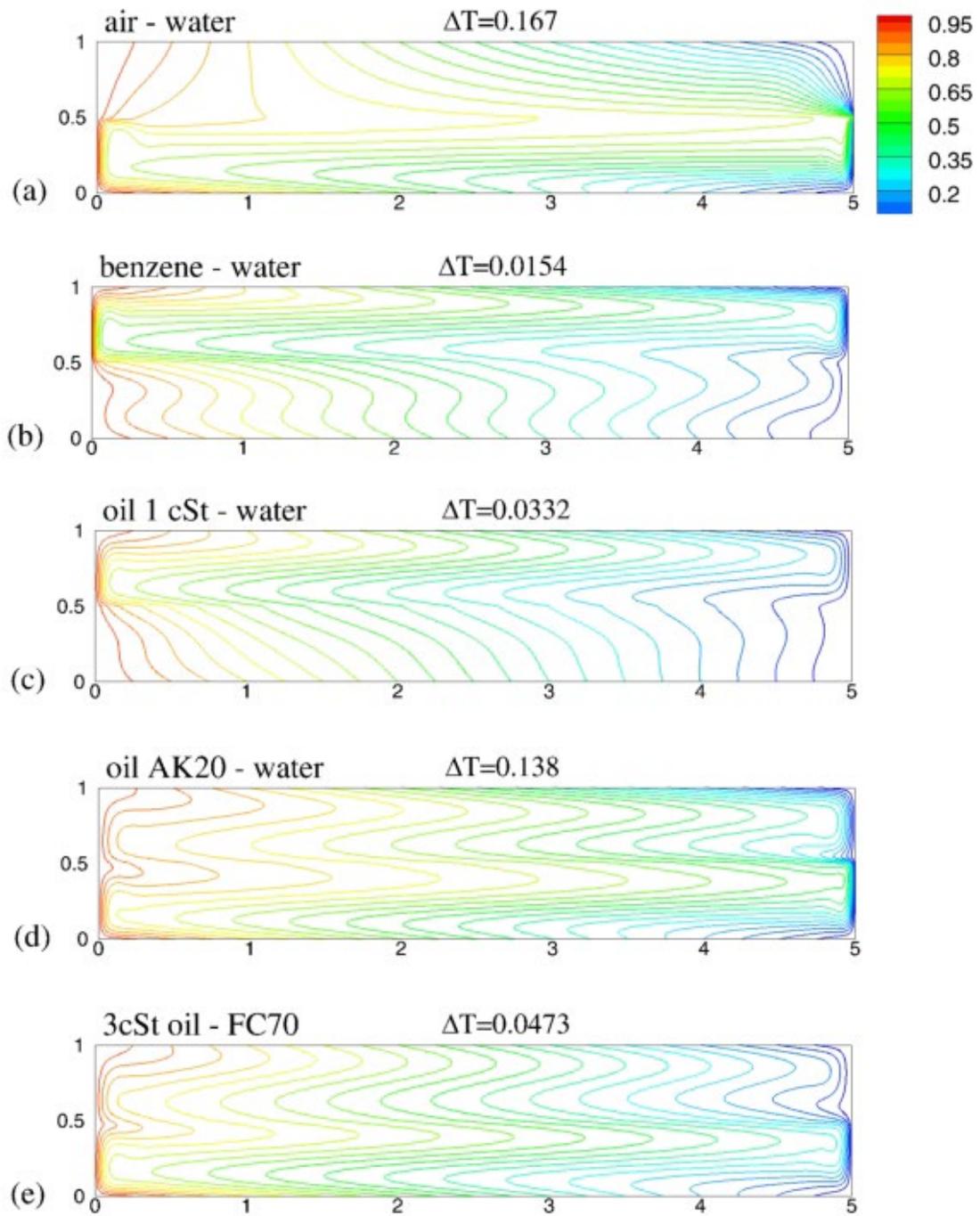

Fig. 3. Isotherms of the considered two fluid systems near their critical points. $H = 10\,cm$, perfectly conducting horizontal boundaries. Uniform temperatures at the vertical boundaries.

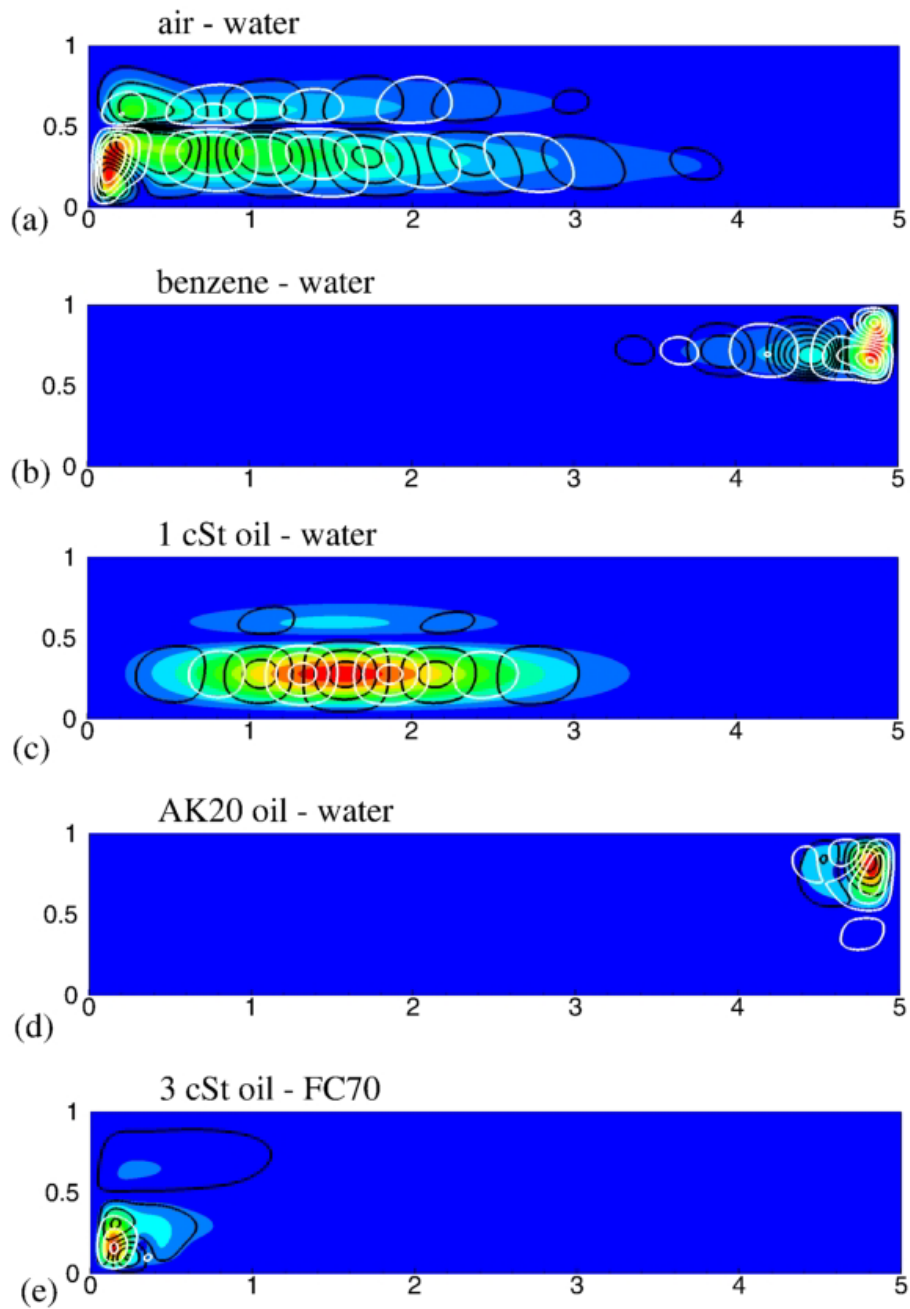

Fig. 4. Most unstable perturbations of the stream function for four cases considered for $H = 10cm$, perfectly conducting horizontal boundaries. The perturbations absolute value is shown by colors, the real and imaginary parts by black and white isolines, respectively. All the levels are equally spaced between the minimal and maximal function values.

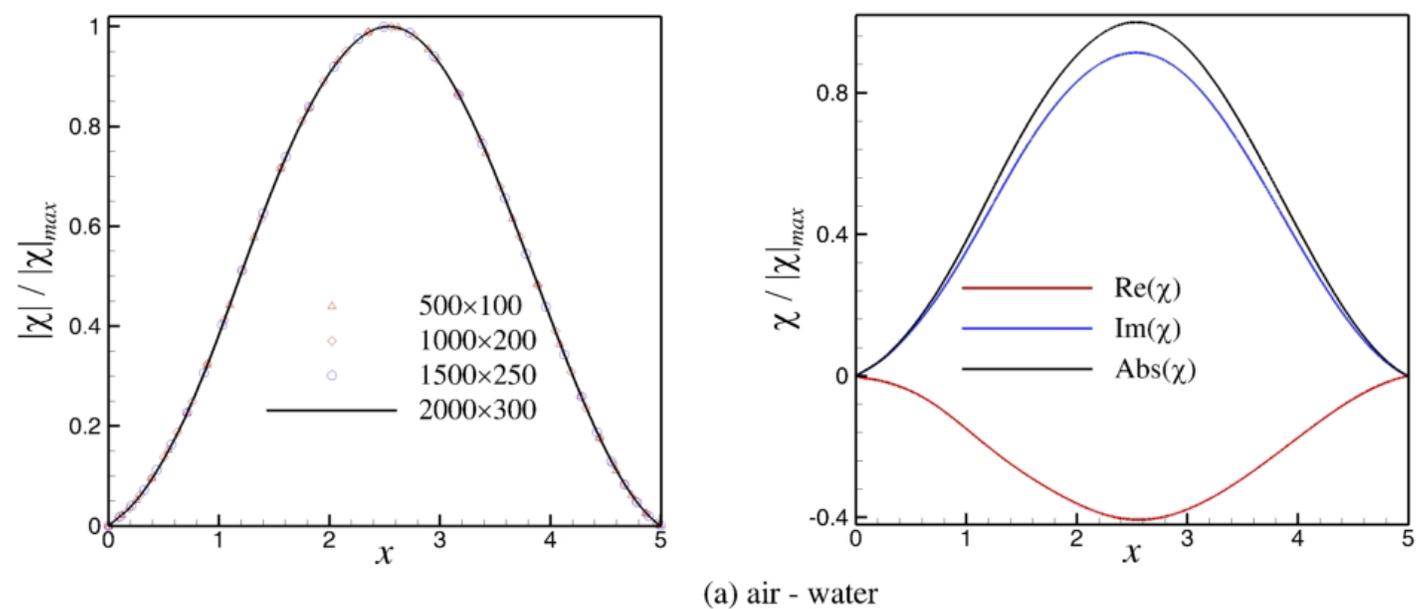
(a) air - water

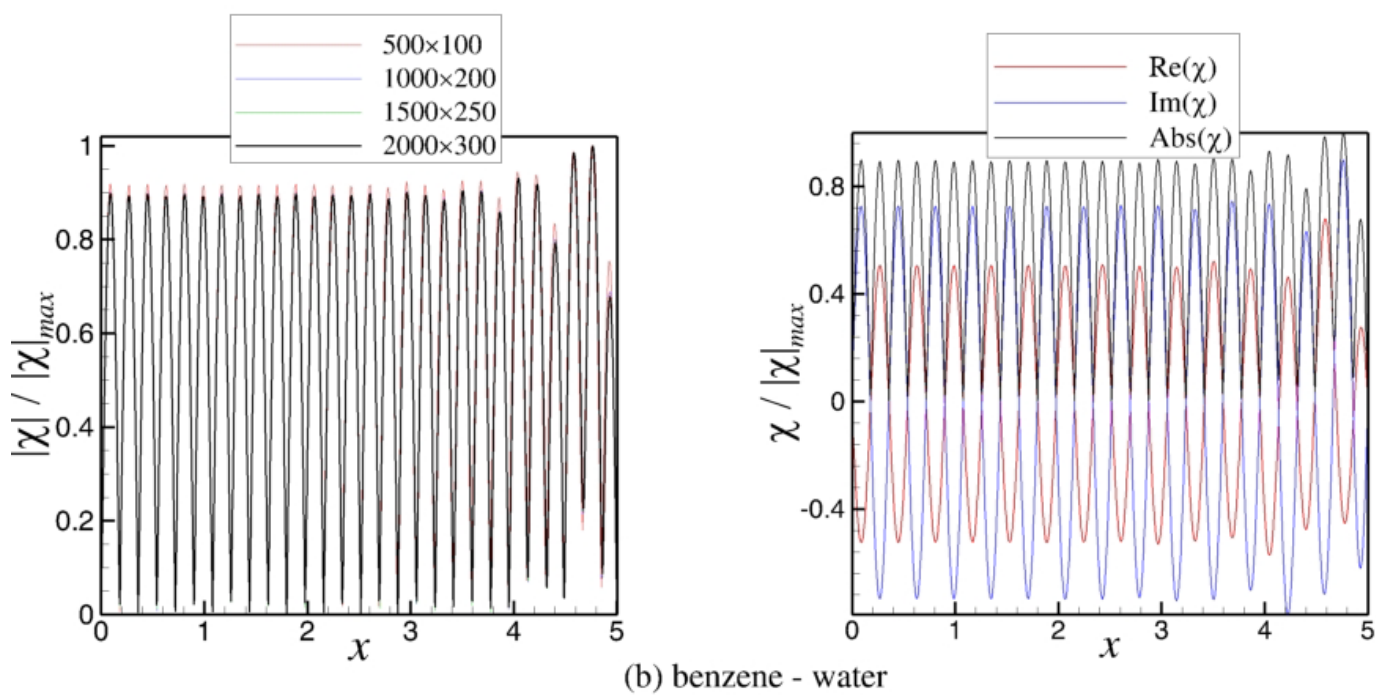
(b) benzene - water

Fig. 5a,b. Disturbances of liquid-liquid interface normalized by their maximal amplitude for $H = 10 cm,$ perfectly conducting horizontal boundaries. Left frames – results for different grids. Right frames – real, imaginary parts and modulus of the disturbances calculated on the finest grid of 2000×300 finite volumes.

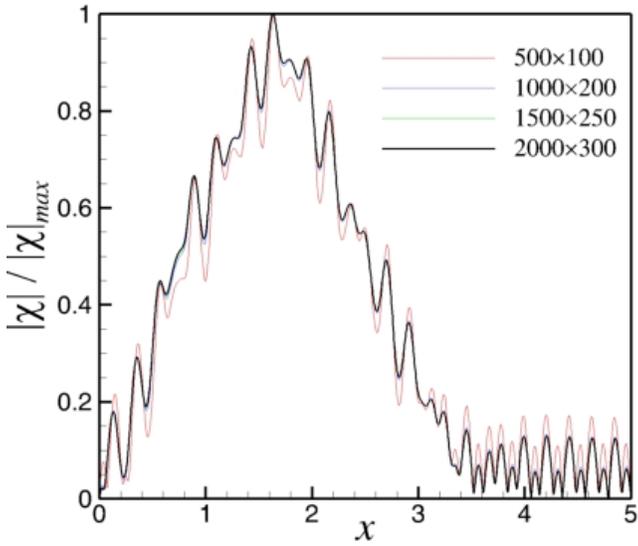
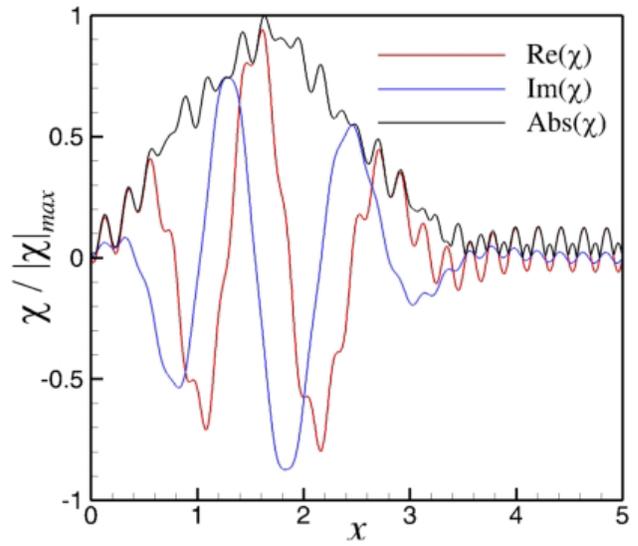

(c) Oil 1cSt - water

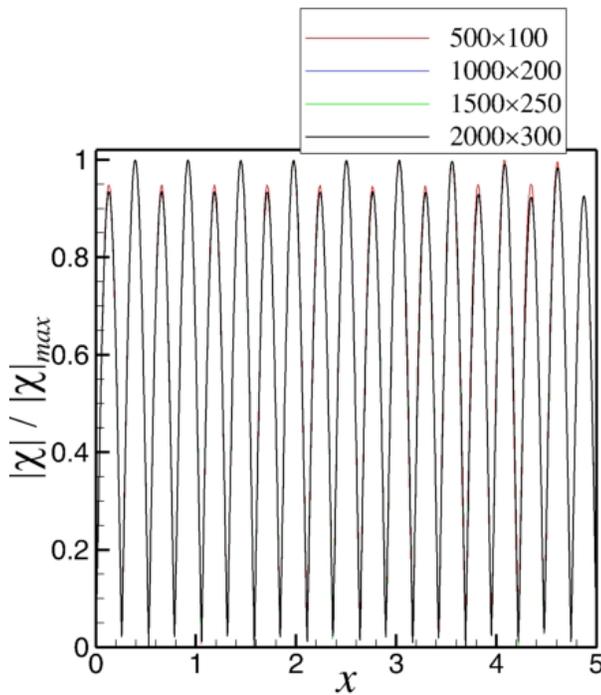
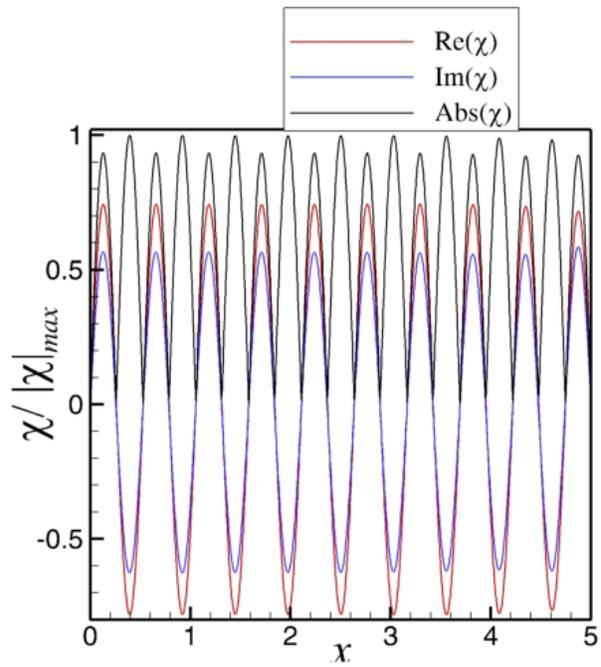

(d) oil AK20 - water

Fig. 5c,d. Disturbances of liquid-liquid interface normalized by their maximal amplitude for $H = 10 cm,$ perfectly conducting horizontal boundaries. Left frames – results for different grids. Right frames – real, imaginary parts and modulus of the disturbances calculated on the finest grid of 2000×300 finite volumes.

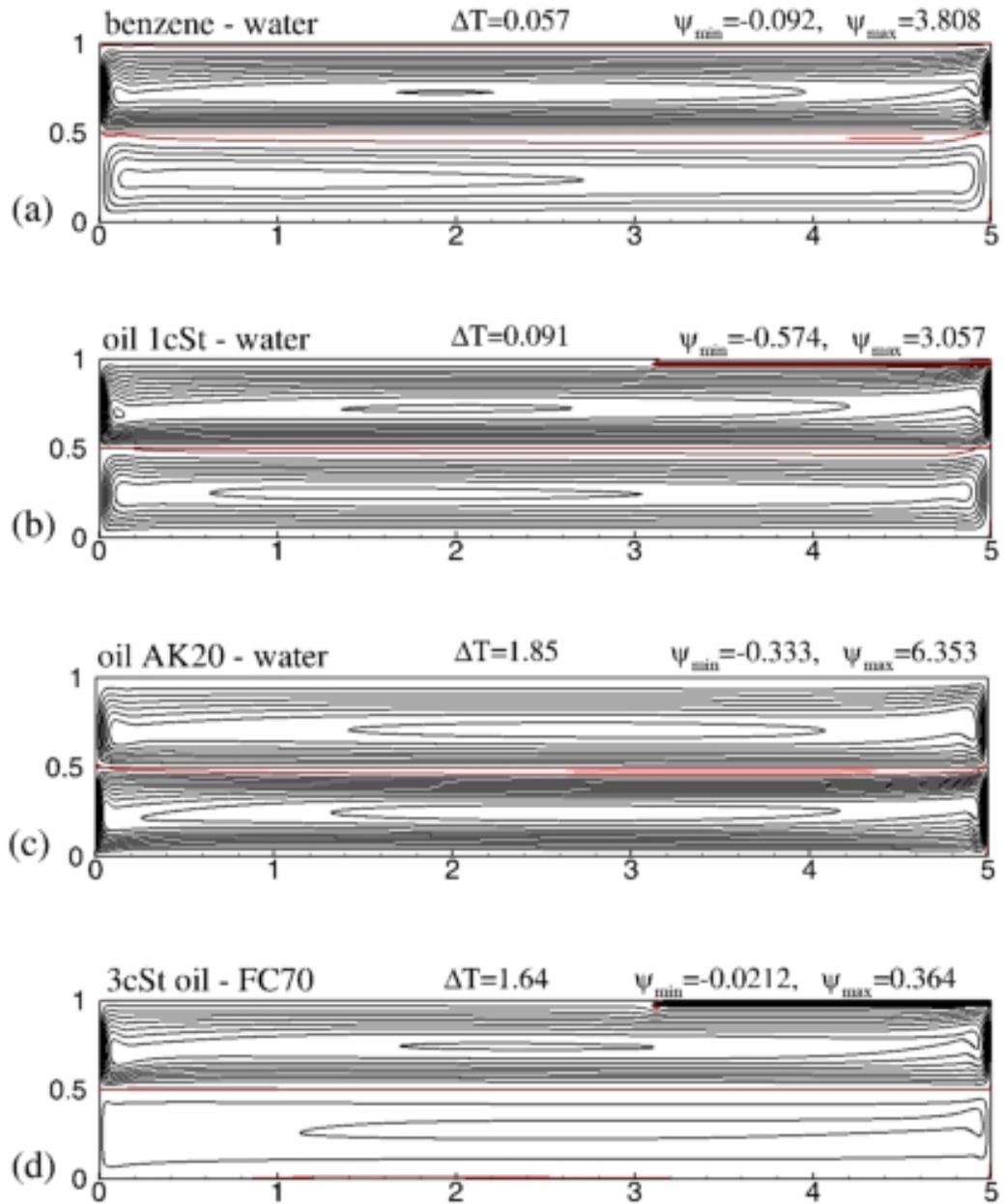

Fig. 6. Streamlines of the considered two fluid systems near their critical points. $H = 10 cm$, perfectly insulating horizontal boundaries. Uniform temperatures at the vertical boundaries. The black streamlines show clockwise circulations, while the red streamlines show counter clockwise circulation.

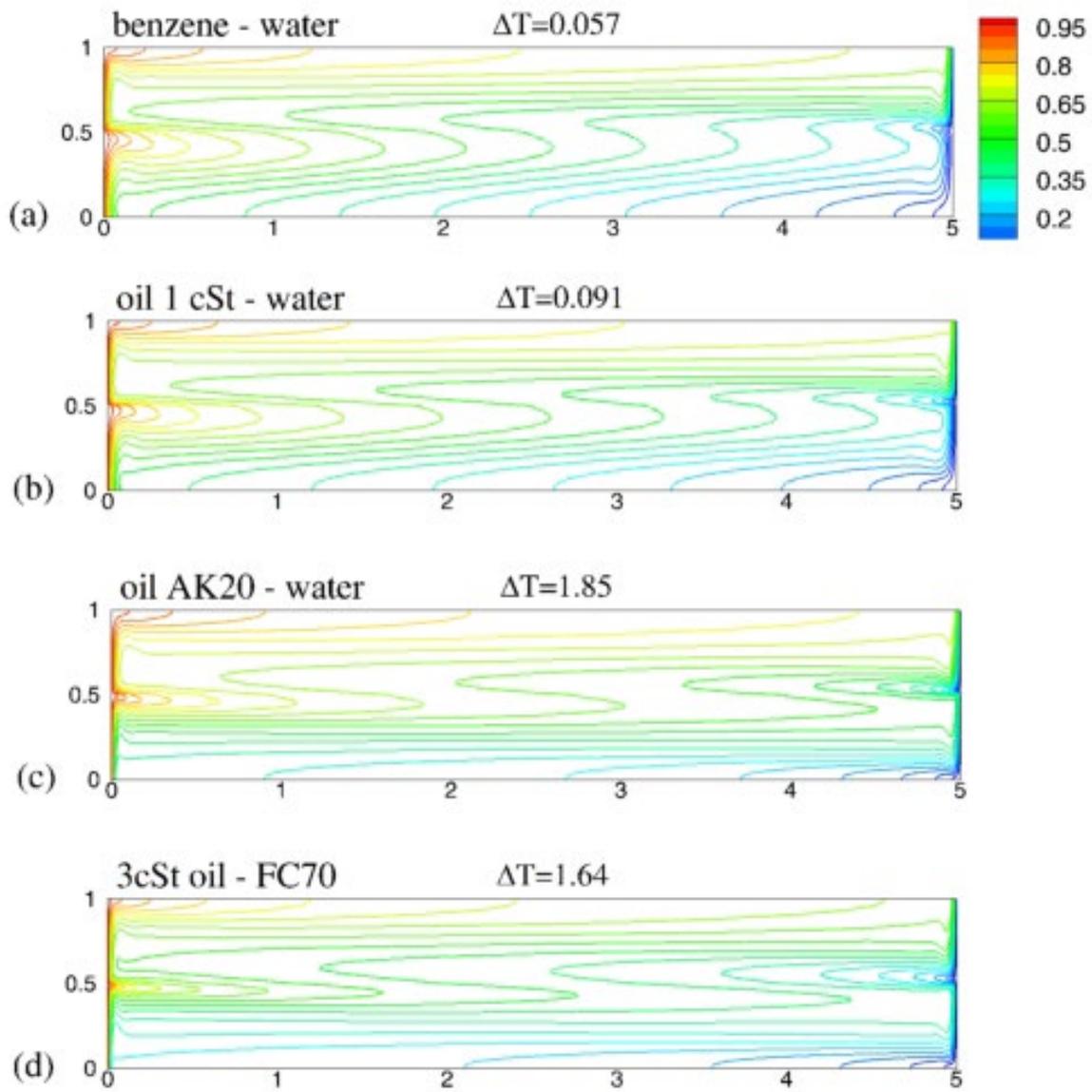

Fig. 7 Isotherms of the considered two fluid systems near their critical points. $H = 10cm$, perfectly insulating horizontal boundaries. Uniform temperatures at the vertical boundaries.

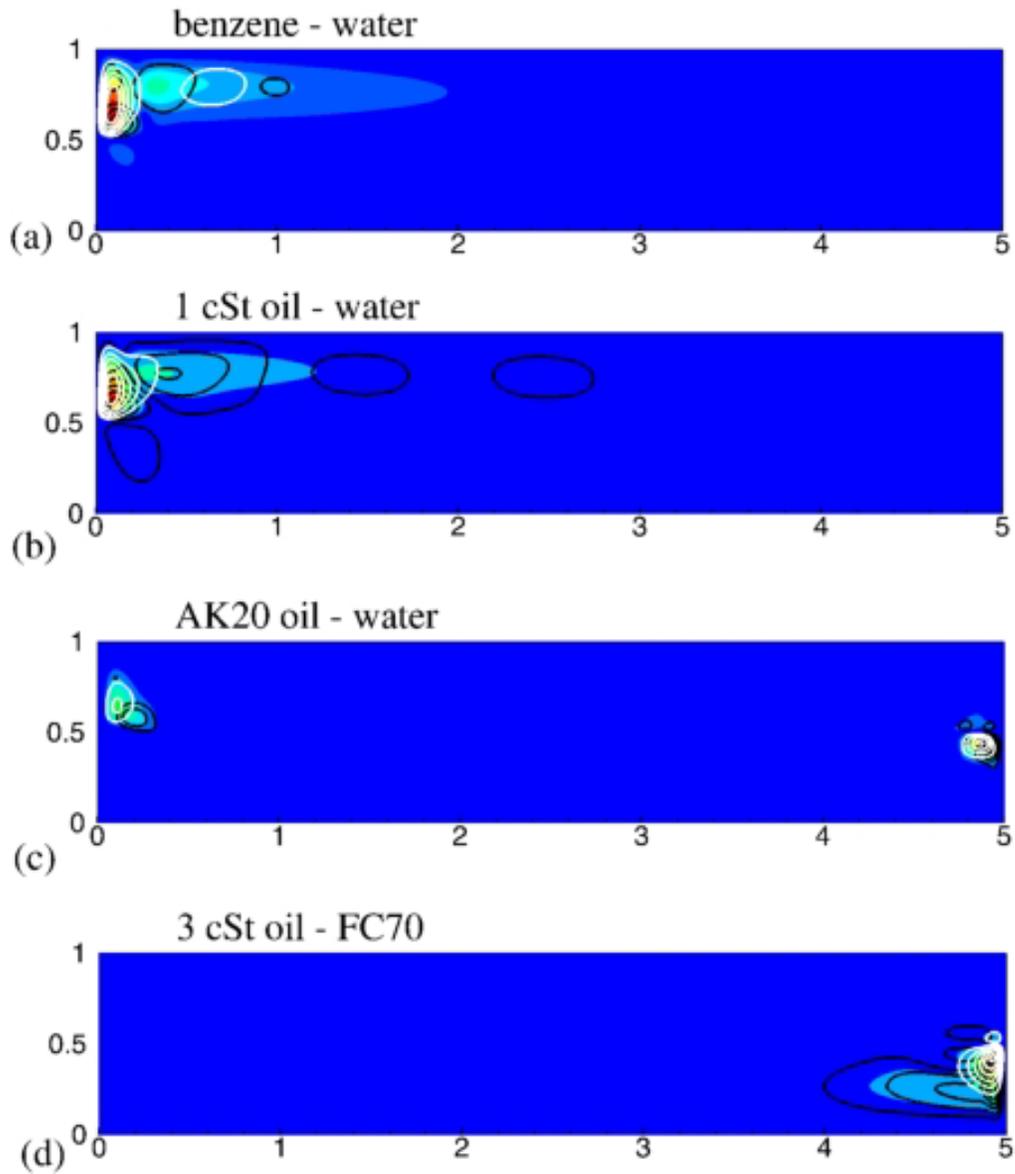

Fig. 8. Most unstable perturbations of the stream function for four cases considered for $H = 10\ cm,$ perfectly insulating horizontal boundaries. The perturbations absolute value is shown by colors, the real and imaginary parts by black and white isolines, respectively. All the levels are equally spaced between the minimal and maximal function values.

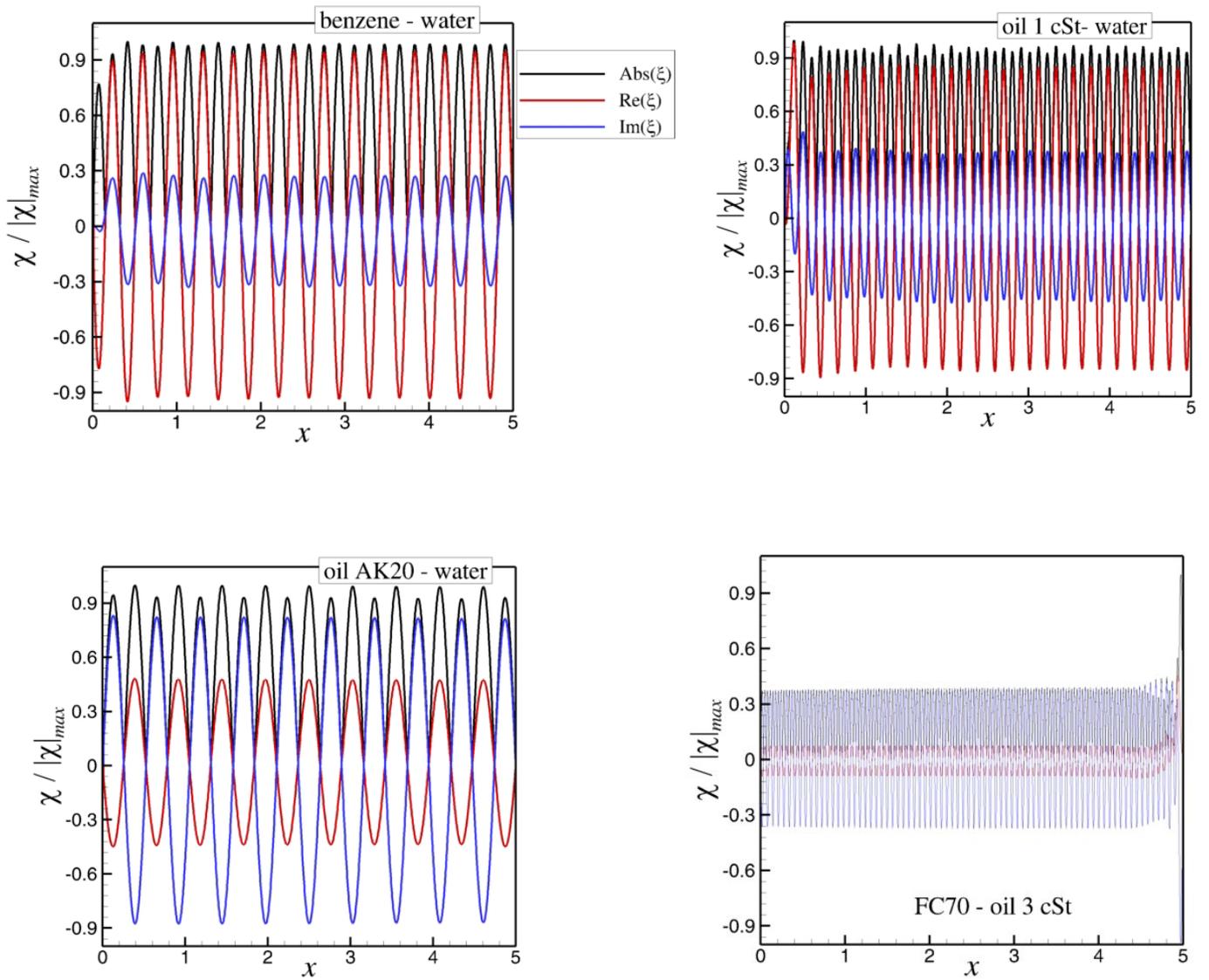

Fig. 9. Disturbances of liquid-liquid interface normalized by their maximal amplitude for $H = 10\ cm,$ perfectly insulating horizontal boundaries. The disturbances calculated on the finest grid of 2000×300 finite volumes.

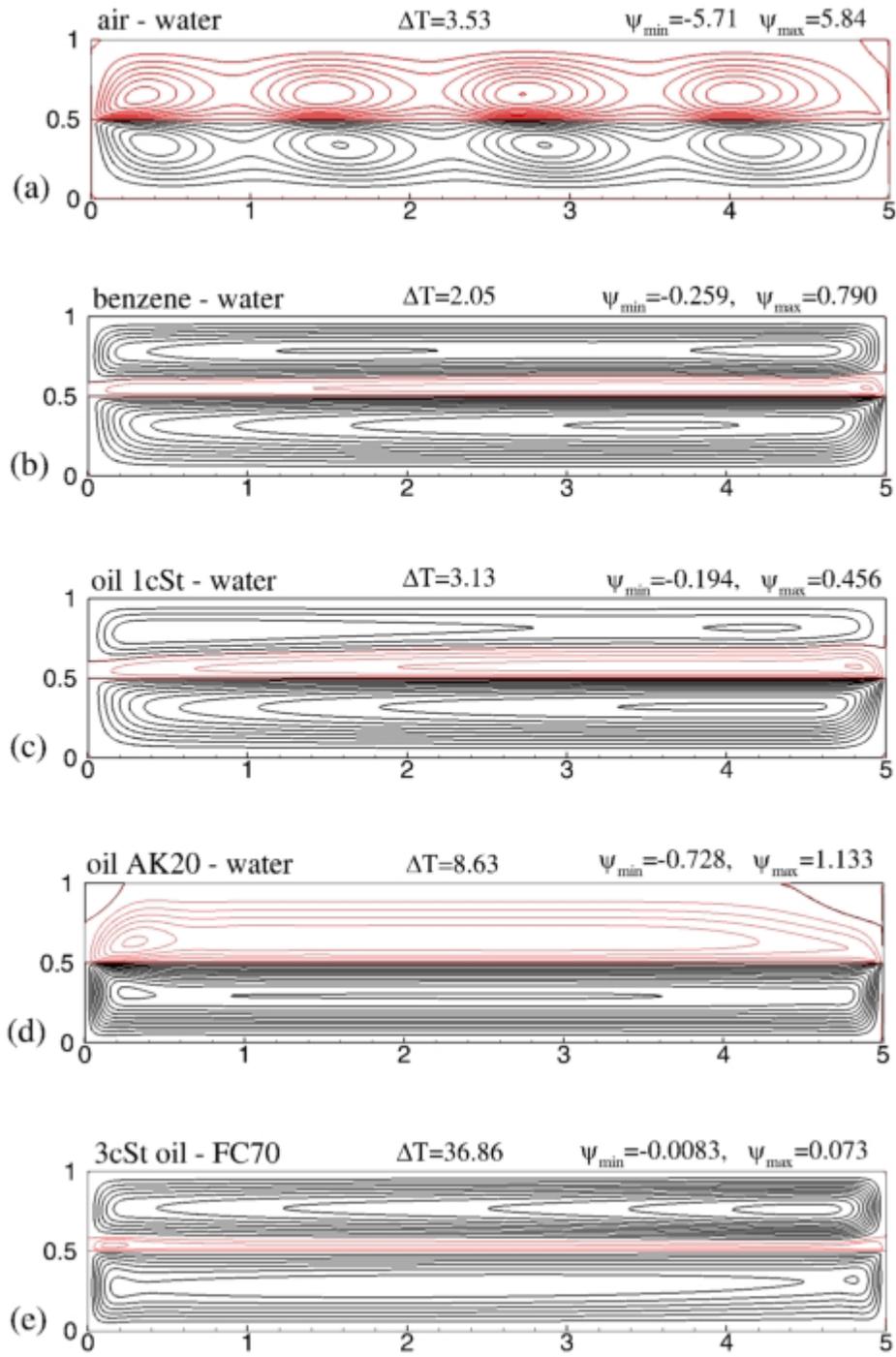

Fig. 10. Streamlines of the considered two fluid systems near their critical points. $H = 1cm$, perfectly conducting horizontal boundaries. Uniform temperatures at the vertical boundaries. The black streamlines show clockwise circulations, while the red streamlines show counter clockwise circulation.

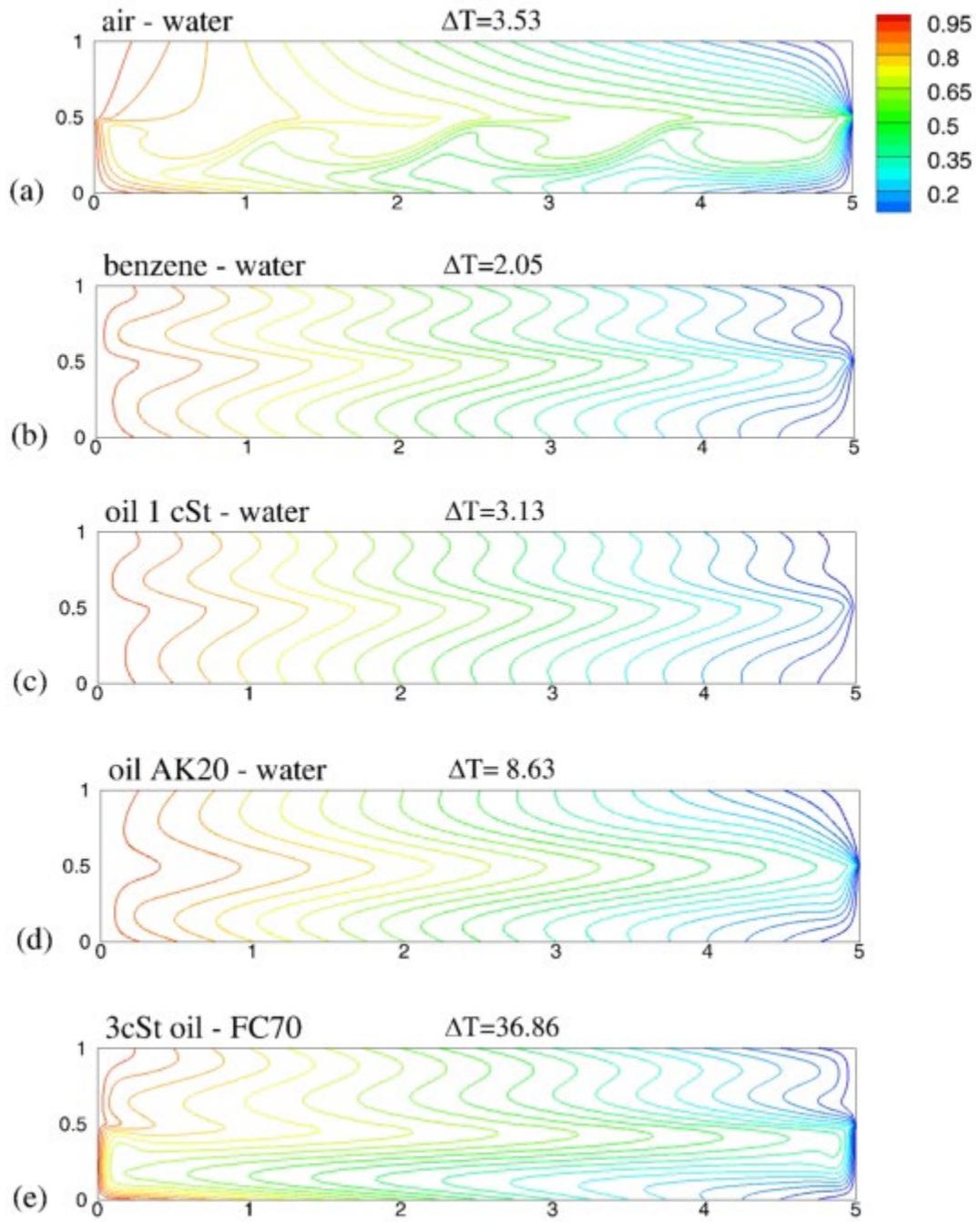

Fig. 11. Isotherms of the considered two fluid systems near their critical points. $H = 1cm$, perfectly conducting horizontal boundaries. Uniform temperatures at the vertical boundaries.

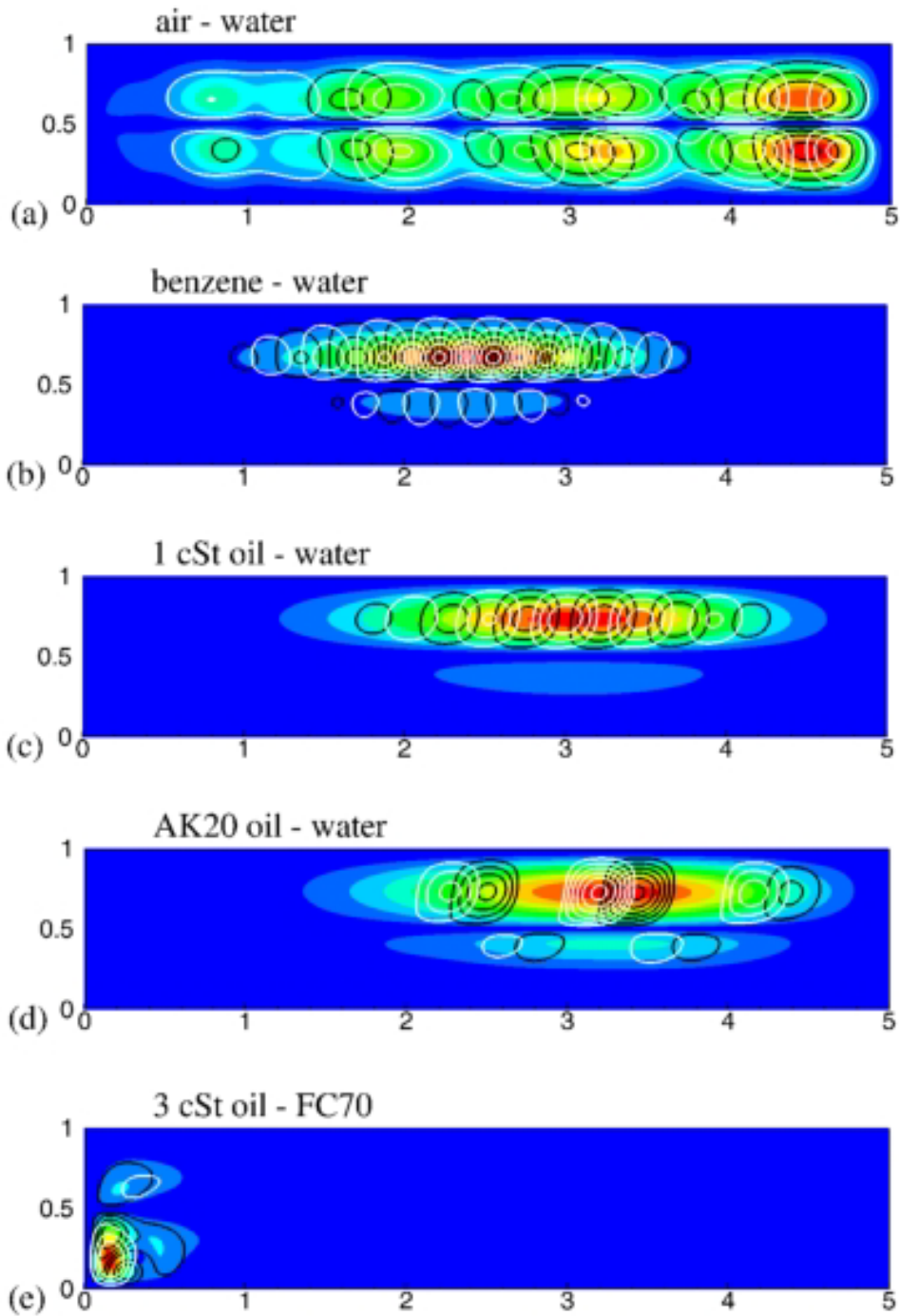

Fig. 12. Most unstable perturbations of the stream function for four cases considered for $H = 1\ cm$, perfectly conducting horizontal boundaries. The perturbations absolute value is shown by colors, the real and imaginary parts by black and white isolines, respectively. All the levels are equally spaced between the minimal and maximal function values.

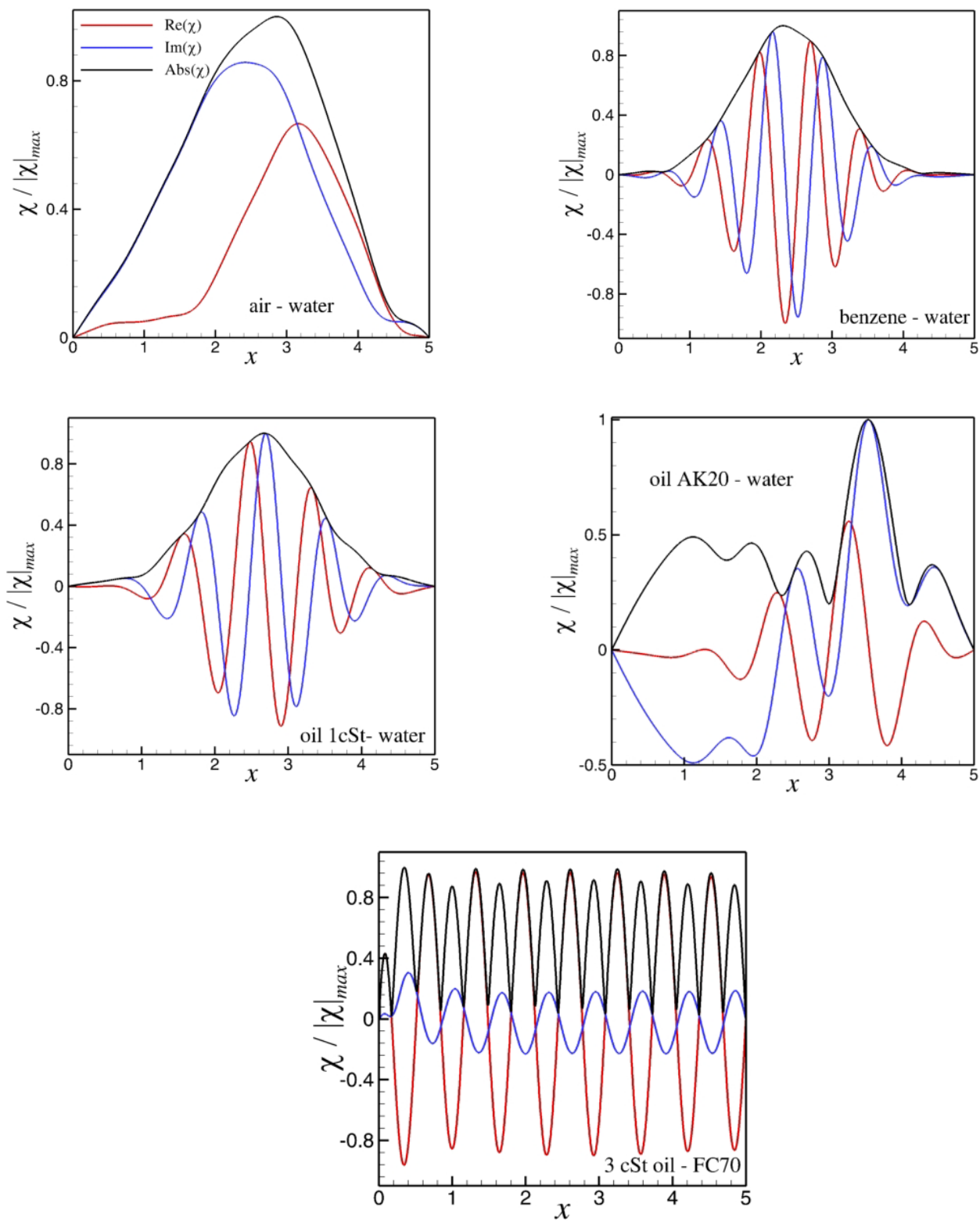

Fig. 13. Disturbances of liquid-liquid interface normalized by their maximal amplitude for $H = 1 cm,$ perfectly conducting horizontal boundaries. The disturbances calculated on the finest grid of 2000×300 finite volumes.

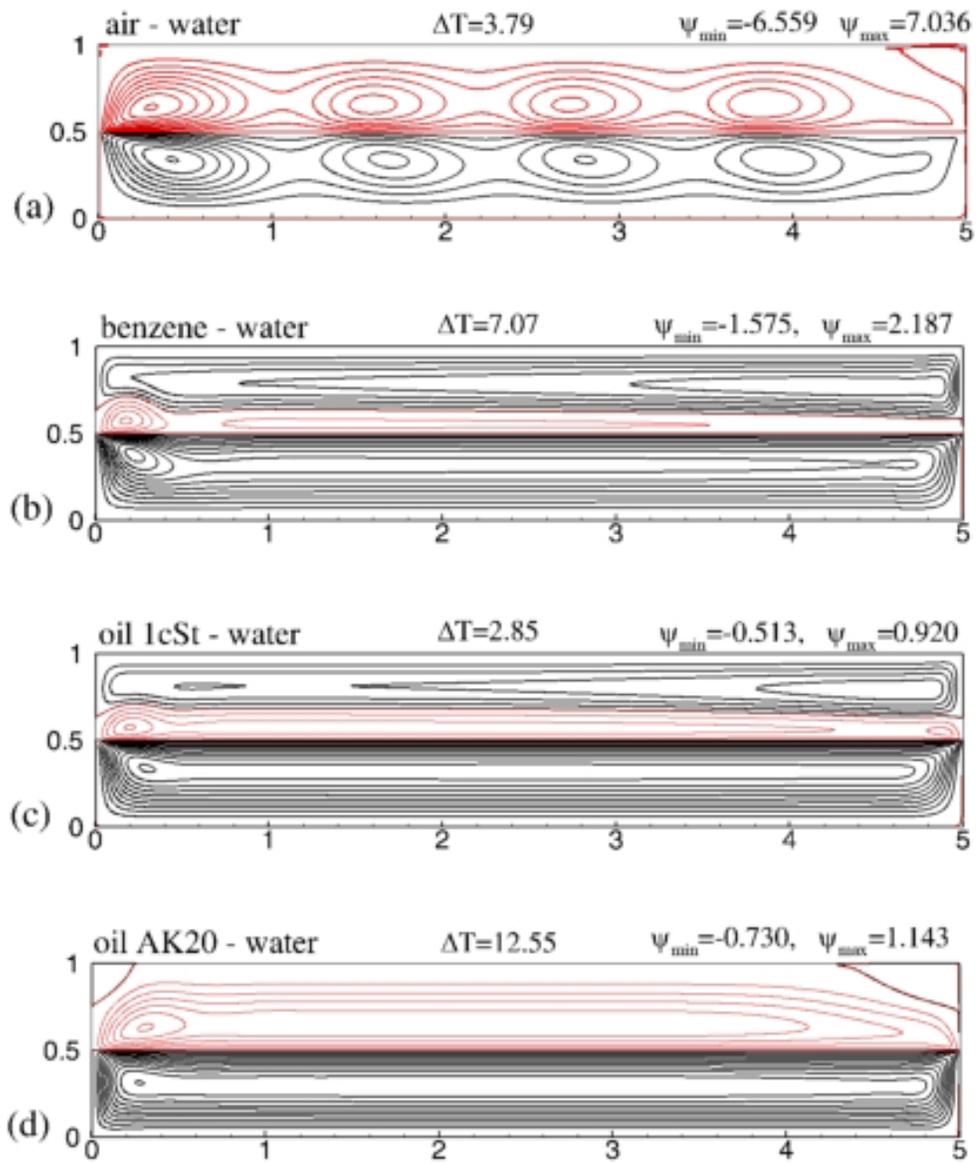

Fig. 14. Streamlines of the considered two fluid systems near their critical points. $H = 1cm,$ perfectly insulating horizontal boundaries. Uniform temperatures at the vertical boundaries. The black streamlines show clockwise circulations, while the red streamlines show counter clockwise circulation.

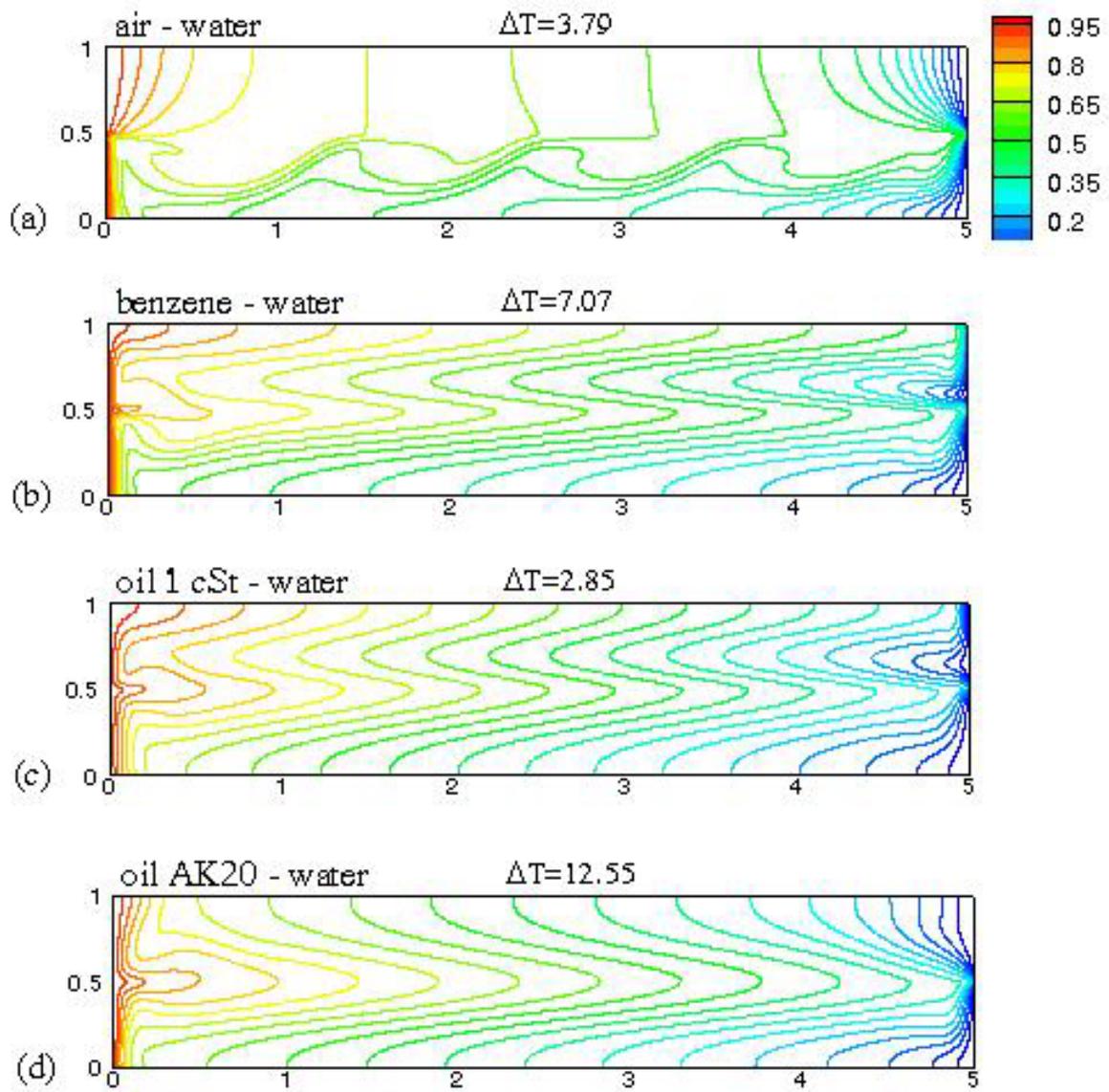

Fig. 15. Isotherms of the considered two fluid systems near their critical points. $H = 1cm$, perfectly insulating horizontal boundaries. Uniform temperatures at the vertical boundaries.

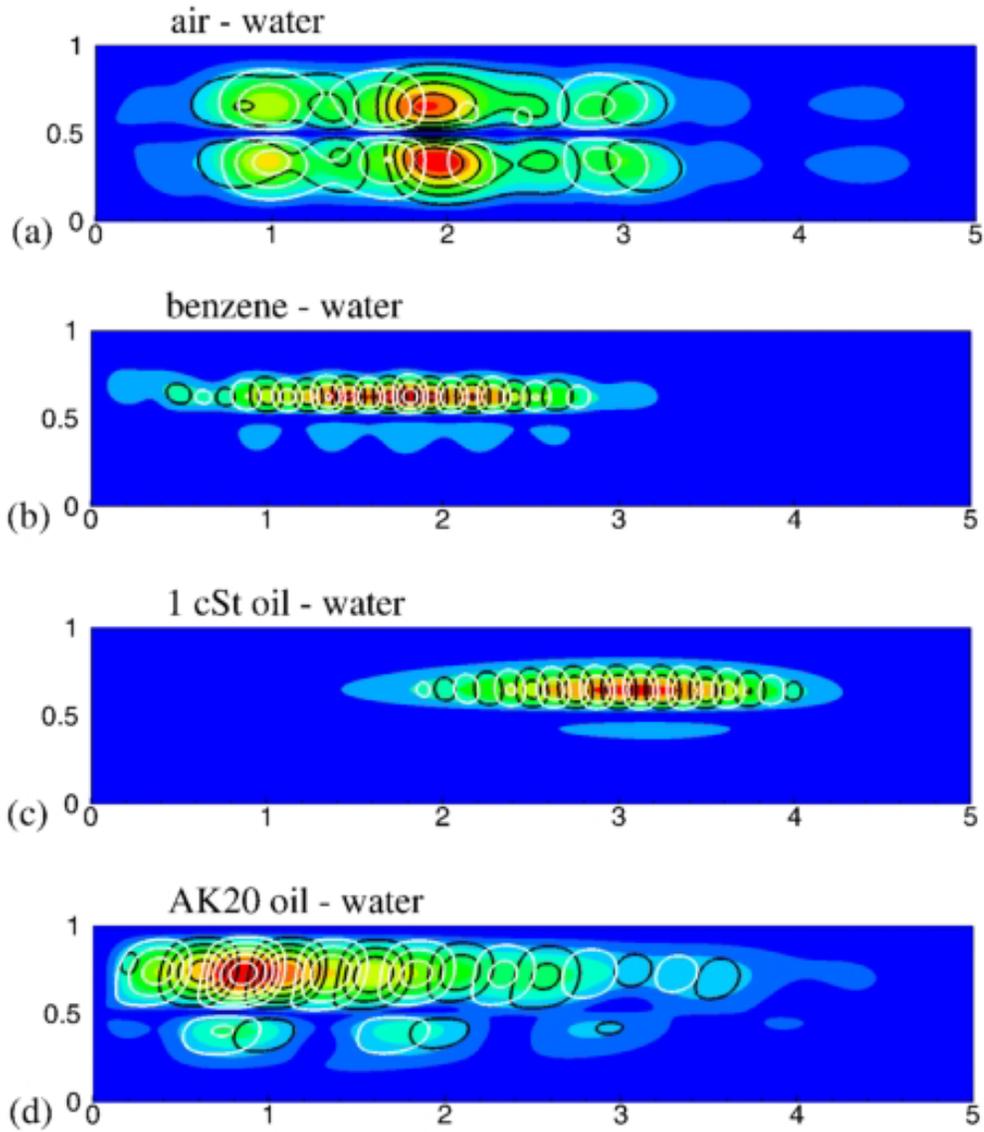

Fig. 16. Most unstable perturbations of the stream function for four cases considered for $H = 1\ cm$, perfectly insulating horizontal boundaries. The perturbations absolute value is shown by colors, the real and imaginary parts by black and white isolines, respectively. All the levels are equally spaced between the minimal and maximal function values.

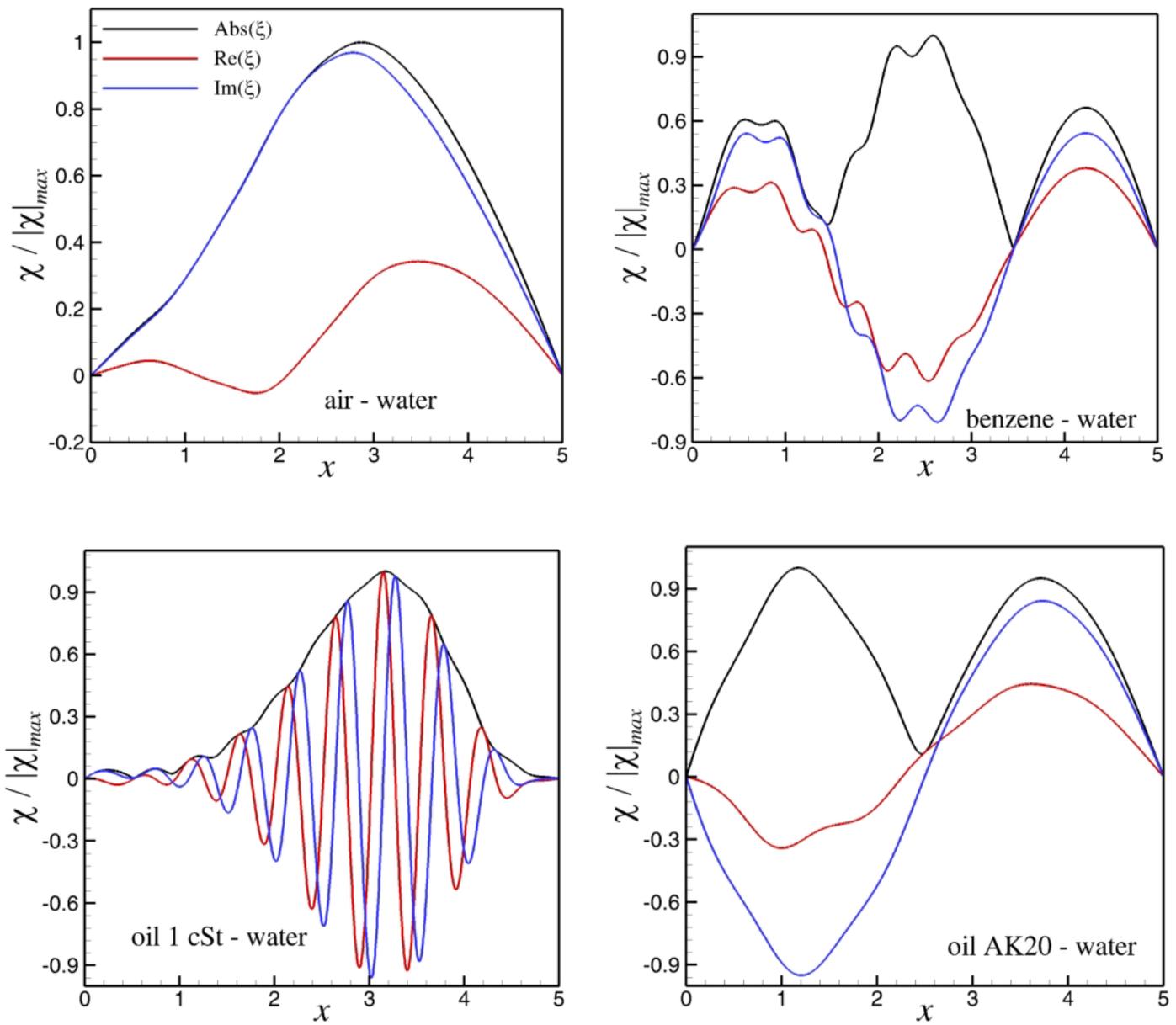

Fig. 17. Disturbances of liquid-liquid interface normalized by their maximal amplitude for $H = 1 cm,$ perfectly insulated horizontal boundaries. The disturbances calculated on the finest grid of 2000×300 finite volumes.

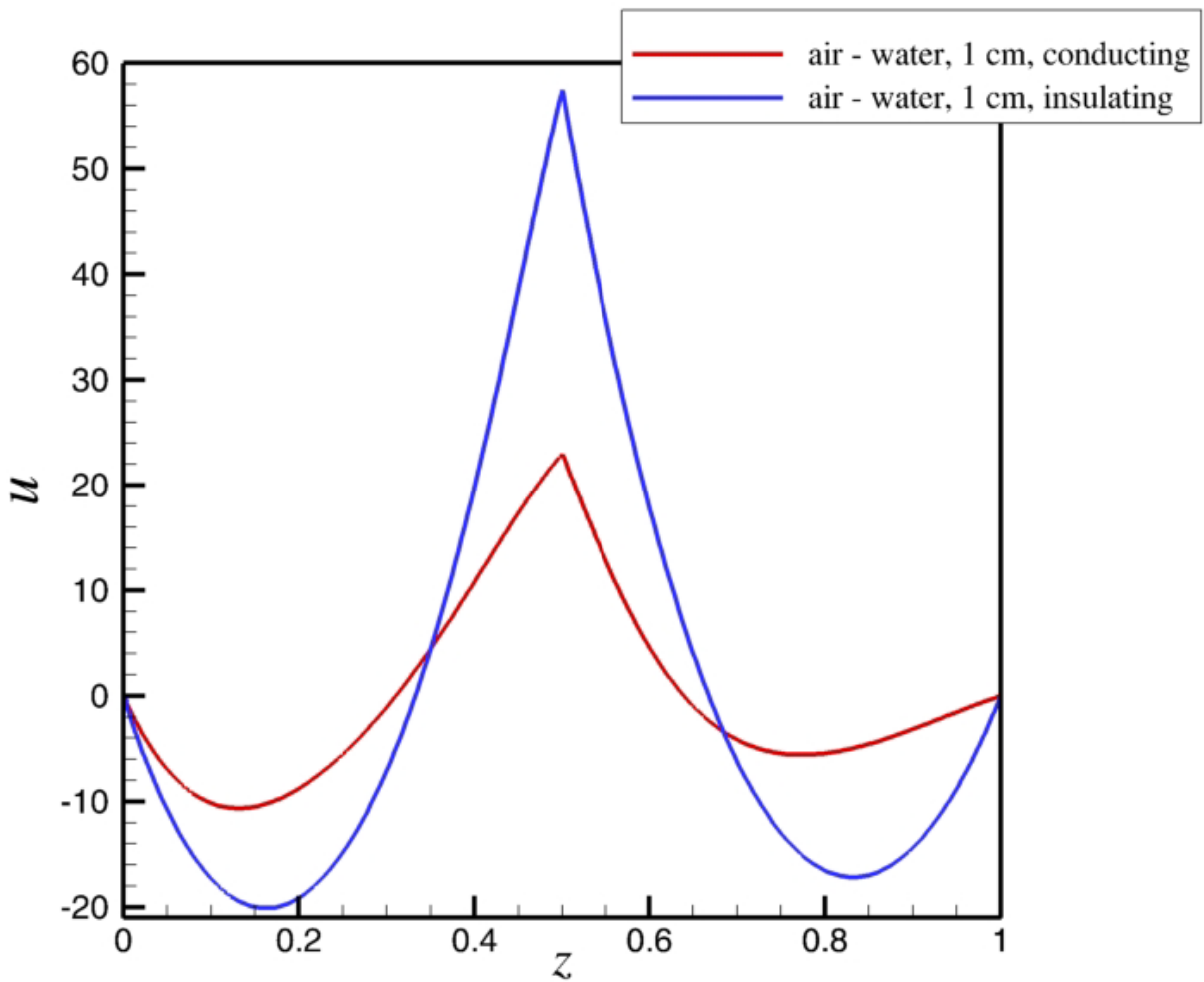

Fig. 19. Vertical profiles of base flow horizontal velocity plotted across the maximum of absolute value of the most unstable perturbation of the horizontal velocity.

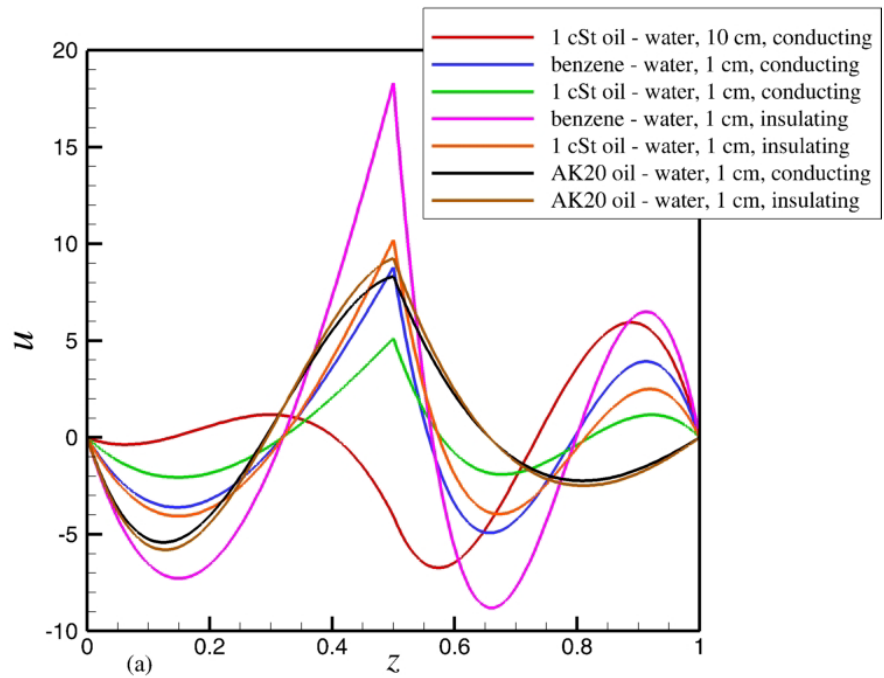 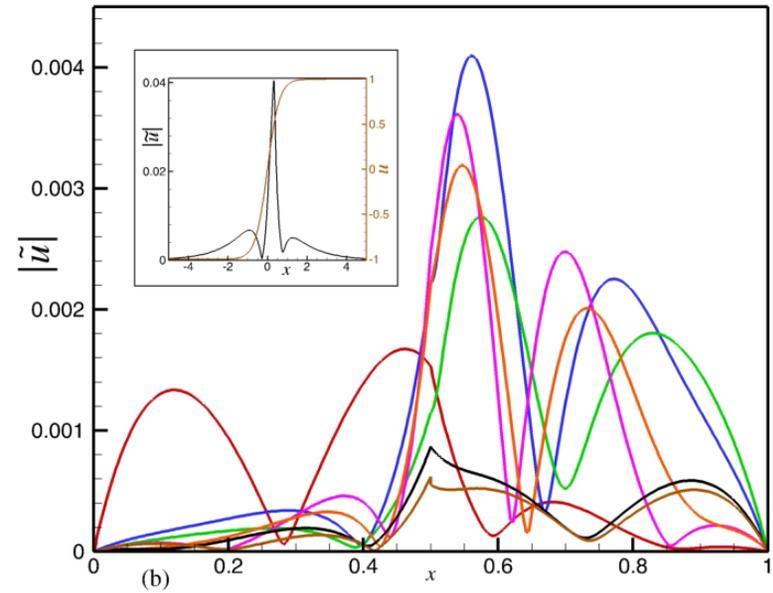

Fig. 19. Velocity profiles having inflection points at the critical temperature differences (a), and profiles of the most unstable perturbation of the horizontal velocity. The profiles are plotted across the maximum of absolute value of the most unstable perturbation of the horizontal velocity. The insert in frame (b) shows velocity profile (brown) and absolute value of perturbation corresponding to the Holmboe instability (black) in a stratified mixing layer (Gelfgat & Kit, 2006).